\documentclass[fleqn,usenatbib]{mnras}
\usepackage{type1cm} % Esto hace que podamos usar los tamaños impuestos por mnras sin problemas.
\usepackage{newtxtext,newtxmath}
\usepackage{amsmath}
\usepackage[utf8]{inputenc}

\usepackage[T1]{fontenc}
\usepackage{ae,aecompl}

\usepackage{graphicx}	% Including figure files
\usepackage{amsmath}	% Advanced maths commands
\usepackage{amssymb}	% Extra maths symbols
\usepackage{placeins}
\usepackage[flushleft]{threeparttable}

\title[High-$z$ LAEs and LBGs SPs]{Differences and similarities of stellar populations in LAEs and LBGs at $\boldsymbol{z\sim}$ 3.4 - 6.8}

\author[P. Arrabal Haro et al.]{P. Arrabal Haro$^{1,2}$\thanks{E-mail: parrabalh@gmail.com}, J. M. Rodríguez Espinosa$^{1,2}$, C. Muñoz-Tuñón$^{1,2}$, D. Sobral$^{3}$,
\newauthor A. Lumbreras-Calle$^{1, 2}$, M. Boquien$^{4}$, A. Hernán-Caballero$^{5}$, L. Rodríguez-Muñoz$^{6}$
\newauthor and B. Alcalde Pampliega$^{7,8}$
\\
$^{1}$Instituto de Astrofísica de Canarias (IAC), E-38205 La Laguna, Spain\\
$^{2}$Departamento de Astrofísica, Universidad de La Laguna, E-38206 La Laguna, Spain\\
$^{3}$Department of Physics, Lancaster University, Lancaster LA1 4YB, UK\\
$^{4}$Centro de Astronomía (CITEVA), Universidad de Antofagasta, Avenida Angamos 601, Antofagasta, Chile\\
$^{5}$Centro de Estudios de Física del Cosmos de Aragón (CEFCA), Plaza San Juan 1, -2, 44001 Teruel, Spain\\
$^{6}$Dipartimento di Fisica e Astronomia, Università di Padova, vicolo dell’Osservatorio 2, I-35122 Padova, Italy\\
$^{7}$Departamento de Física de la Tierra y Astrofísica, Faultad de CC Físicas, Universidad Complutense de Madrid, E-2840 Madrid, Spain\\
$^{8}$Isaac Newton Group of Telescopes (ING), Apto. 321, E-38700 Santa Cruz de la Palma, Canary Islands, Spain
}

\date{Accepted 2020 April 21. Received 2020 April 3; in original form 2019 August 14}

\pubyear{2019}

\begin{document}
\label{firstpage}
\pagerange{\pageref{firstpage}--\pageref{lastpage}}
\maketitle

% Abstract of the paper. It should be a single paragraph not more than 250 words (200 words for Letters). No references should appear in the abstract.
\begin{abstract}
LAEs and LBGs represent the most common groups of star-forming galaxies at \mbox{high-$z$}, and the differences between their inherent stellar populations (SPs) are a key factor in understanding early galaxy formation and evolution. We have run a set of SP \mbox{burst-like} models for a sample of 1,558 sources at \mbox{$3.4<z<6.8$} from the Survey for \mbox{High-$z$} Absorption Red and Dead Sources (SHARDS) over the GOODS-N field. This work focuses on the differences between the three different observational subfamilies of our sample: \mbox{LAE-LBGs}, \mbox{no-Ly$\alpha$} LBGs and pure LAEs. Single and double SP synthetic spectra were used to model the SEDs, adopting a Bayesian information criterion to analyse under which situations a second SP is required. We find that the sources are well modelled using a single SP in \mbox{$\sim79\%$} of the cases. The best models suggest that pure LAEs are typically young low mass galaxies (\mbox{$t\sim26^{+41}_{-25}$ Myr}; \mbox{$M_{\mathrm{star}}\sim5.6^{+12.0}_{-5.5}\times10^{8}\ M_{\odot}$}), undergoing one of their first bursts of star formation. On the other hand, \mbox{no-Ly$\alpha$} LBGs require older SPs (\mbox{$t\sim71\pm12$ Myr}), and they are substantially more massive (\mbox{$M_{\mathrm{star}}\sim3.5\pm1.1\times10^{9}\ M_{\odot}$}). \mbox{LAE-LBGs} appear as the subgroup that more frequently needs the addition of a second SP, representing an old and massive galaxy caught in a strong recent star-forming episode. The relative number of sources found from each subfamily at each $z$ supports an evolutionary scenario from pure LAEs and single SP \mbox{LAE-LBGs} to more massive LBGs. Stellar Mass Functions are also derived, finding an increase of $M^{*}$ with cosmic time and a possible steepening of the low mass slope from \mbox{$z\sim6$} to \mbox{$z\sim5$} with no significant change to \mbox{$z\sim4$}. Additionally, we have derived the \mbox{SFR-$M_{\mathrm{star}}$} relation, finding a \mbox{$\mathrm{SFR}\propto M_{\mathrm{star}}^{\beta}$} behaviour with negligible evolution from \mbox{$z\sim4$} to \mbox{$z\sim6$}.
\end{abstract}

% Select between one and six entries from the list of approved keywords.
% Don't make up new ones.

%%%%%%%%%%%%%%%%%%%%%%%%%%%%%%%%%%%%%%%%%%%%%%%%%%

%%%%%%%%%%%%%%%%% BODY OF PAPER %%%%%%%%%%%%%%%%%%

\section{Introduction}
\label{sec:Introduction}
Lyman Alpha Emitters (LAEs) and Lyman Break Galaxies (LBGs) have traditionally been the two main types of \mbox{high-$z$} star-forming galaxies. They are typically detected in the optical and Near Infrared (NIR) through their redshifted Ly$\alpha$ line and Lyman continuum break \citep[\textit{e.g.},][]{Koo1980, Steidel1993, Giavalisco1996, Ouchi2009, Robertson2010, Bouwens2011, Matthee2017, Sobral2018}. The usual separation into these two families is due to the selection techniques involved in their detection, as well as the presence or not of the Ly$\alpha$ emission line at high Equivalent Width (EW). LAEs have traditionally been detected using narrow band filters \citep[][among others]{Hu1998,Malhotra2004,Taniguchi2005,Iye2006,Gronwall2007,Ouchi2008,Ouchi2010,Cassata2015,Santos2016, Matthee2017, Sobral2017,Sobral2018}. This technique usually employs complementary broad-band filters. The comparison of the emissions detected in the narrow-band with the broad band sampling a similar wavelength makes it possible to identify emission excesses in the narrow band corresponding to the Ly$\alpha$ line emission. On the other side, deep broad band images have been typically used to detect LBGs through the Lyman-break technique \citep[as in, \textit{e.g.},][among others]{Steidel2003,Giavalisco2004,Iwata2007,McLure2009,Oesch2010,vanderBurg2010,Ellis2013,Bouwens2014, Bouwens2015, Laporte2016}. Some authors have modelled LAEs and compared them with the LBGs, claiming that LAEs represent a less luminous LBG subset. Other works conclude that LAEs and LBGs are essentially similar, the difference being solely in the technique involved in their detection \citep{Dayal2012}. However, other authors \citep[\textit{e.g.},][]{Giavalisco2002, Gawiser2006} claim that LAEs are low mass sources, with little dust and rapid star formation. In any case, the lack of sufficient spectroscopy of sources at \mbox{high-$z$} has maintained the usual separation normally assumed between LAEs and LBGs.

Traditional narrow and broad band detection techniques could imply missing the ultraviolet (UV) continuum in LAEs or getting the LBGs line emission diluted in the broad band filters, hence the advantage of employing a large set of multiple consecutive medium/narrow filters to better identify emission lines, as done in \textit{e.g.}, \citet{RodriguezEspinosa2014}, \citet{Cava2015}, \citet{HernanCaballero2017}, \citet{ArrabalHaro2018} and \citet{LumbrerasCalle2019} to detect line emitters of different nature in the Survey for \mbox{High-$z$} Absorption Red and Dead Sources \citep[SHARDS,][]{PerezGonzalez2013}. A filter configuration of that characteristics not only provides Spectral Energy Distributions (SEDs) with better spectral resolution, which play a key role in the rejection of lower redshift interlopers, as shown in \citet{ArrabalHaro2018}, but it also allows to select LAEs and LBGs simultaneously in a systematic way, as achieved with MUSE (Multi Unit Spectroscopic Explorer) in \citet{Bina2016}, \citet{Drake2017} or \citet{Drake2017b}. In this work, we make use of the SHARDS survey, which covers the GOODS-N field in the wavelength range between \mbox{500-941} nm, with a set of 25 consecutive medium band filters, thereby allowing the detection of both LAEs and LBGs from \mbox{$z\sim3.4$} to \mbox{$z\sim6.8$}, as shown in \citet{ArrabalHaro2018}.

Throughout this study we follow the definition given in \textit{e.g.}, \citet{Iye2011}, where any galaxy with Ly$\alpha$ emission line is a LAE. This applies to sources with rest-frame Ly$\alpha$ EW above 5.1 \AA{} in our sample \citep[see][]{ArrabalHaro2018}. The term LBG is reserved for galaxies showing the Lyman break and a well detected rest-frame UV continuum at redder wavelengths. Note that by definition an object can simultaneously be a LAE and a LBG. The sources that present Ly$\alpha$ line emission on top of a well defined rest-frame UV continuum are named \mbox{LAE-LBGs}.

In fact, all LAEs should be LBGs. However, many LAEs can be so faint that their continuum is not detected with the Lyman break dropout technique \citep{Trainor2015, Trainor2016}. We will observationally call ``pure LAEs" to those emitters with a prominent Ly$\alpha$ line and a very faint UV continuum not detected in SHARDS (\mbox{$m_{1500}\gtrsim27.0$ AB}). The term ``\mbox{no-Ly$\alpha$} LBG" will be used for those LBGs exclusively selected through their Lyman break and not presenting Ly$\alpha$ emission line up to our observational limit (Ly$\alpha$ \mbox{EW$_{0}\lesssim5.1$ \AA{}}).

We present herein the results of stellar population (SP) synthesis models fitted to the SEDs of a sample of 1,558 \mbox{high-$z$} galaxies. We pay special attention to whether or not two separated SPs are needed to model the various types of sources. We estimate the age and $M_{\mathrm{star}}$ differences between the observational classes, as well as their relative proportion with redshift. The paper is structured as follows: Section~\ref{sec:Data} gives a quick overview of the sample previously selected and the photometric data employed; Section~\ref{sec:Methods} describes the simulations and the criteria followed to decide between single or double SP; Section~\ref{sec:Results} presents the results; Section~\ref{sec:Discussion} discusses the main physical parameters derived from the models as well as the relation between pure LAEs and LBGs; Section~\ref{sec:Conclusions} summarises the main conclusions. All calculations are made adopting a \mbox{$\Lambda$-dominated} flat universe with \mbox{$H_{0}=68$ km s$^{-1}$ Mpc$^{-1}$}, \mbox{$\Omega_{M}=0.3$} and \mbox{$\Omega_{\Lambda}=0.7$} \citep{Planck2016} and a \citet{Salpeter1955} IMF. All magnitudes are expressed in the AB system \citep{Oke1983}.

\section{Working Data}
\label{sec:Data}
\citet{ArrabalHaro2018} used the 25 \mbox{medium-width} filters \mbox{(FWHM $\sim17$ nm)} of the SHARDS ESO/GTC survey \citep{PerezGonzalez2013} to simultaneously select LAEs and LBGs. The sample of \mbox{high-$z$} galaxies was selected via colour excesses and photometric fits of their SEDs. A complete discussion of the sample build up, as well as the ancillary GOODS-N data used, can be found in \citet{ArrabalHaro2018}, where the coordinates, redshifts, rest-frame Ly$\alpha$ EWs, SFRs, Luminosity Functions (LFs), and other physical parameters are given.

\begin{table}
    \centering
     \caption{Sample distribution among the three different observationally defined subfamilies.}
    \label{tab:sample_EW}
    \begin{threeparttable}
    \begin{tabular}{lccc}
    \hline
        Type    & Defining observational criteria &   N\\
\hline
  No-Ly$\alpha$ LBGs   & $m_{1500}\lesssim27$ AB; $\ \mathrm{EW}_{\mathrm{Ly}\alpha}\lesssim5$ \AA{} & 1030\\
  LAE-LBGs         & $m_{1500}\lesssim27$ AB; $\ \mathrm{EW}_{\mathrm{Ly}\alpha}\gtrsim5$ \AA{} & 404\\
  Pure LAEs         & $\ m_{1500}\gtrsim27$ AB; $\ \mathrm{EW}_{\mathrm{Ly}\alpha}\gtrsim5$ \AA{}$\tnote{1}$ & 124\\
  \hline
    \end{tabular}
        \begin{tablenotes}
     \item[1] Even though this was the original Ly$\alpha$ rest-frame EW criterion, all pure LAEs presented values above 35 \AA{}.
    \end{tablenotes}
    \end{threeparttable}
\end{table}

\begin{figure}
	\includegraphics[width=\columnwidth]{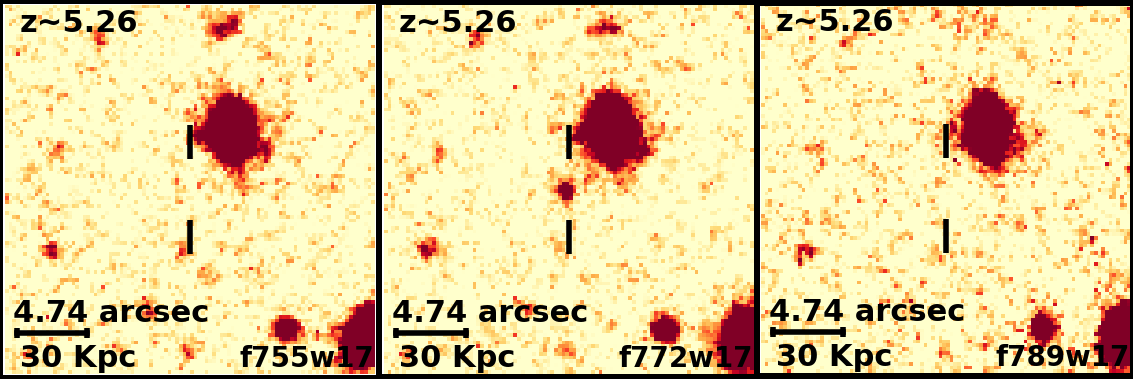}
   \caption{Mosaic of three consecutive SHARDS filters sampling the Ly$\alpha$ emission of the \mbox{$z\sim5.26$} pure LAE SHARDS J123720.02+621200.6 (within the vertical marks). The central filter of the image shows the Ly$\alpha$ line in emission. There is no detection at shorter wavelengths, but neither \mbox{red-ward} of the central filter, since the UV continuum is below the SHARDS detection limit. This source shows a weak continuum detection in \textit{HST}/ACS images \mbox{red-ward} of Ly$\alpha$, though we use the SHARDS images as reference for our definition of pure LAEs. North is up, East is left.}
    \label{fig:Pure_LAE}
\end{figure}

The final sample consists of 1,558 sources at \mbox{$z\sim3.4$-6.8}, distributed into 1,434 LBGs (404 of them showing Ly$\alpha$ emission line with \mbox{$\mathrm{EW_{0}}>5.1$ \AA{}}), and 124 pure LAEs (\mbox{$m_{1500}\gtrsim27.0$ AB}; Ly$\alpha$ \mbox{$\mathrm{EW_{0}}>35$ \AA{}}) as summarised in Table~\ref{tab:sample_EW}. Note that pure LAEs were originally selected as faint continuum sources with a prominent emission in one of the SHARDS filters representing the Ly$\alpha$ line. Because of this, all of them present \mbox{$\mathrm{EW_{0}}>35$ \AA{}} \citep[see][]{ArrabalHaro2018}. An example of a pure LAE is shown in Fig.~\ref{fig:Pure_LAE}. In order to further extend our SEDs beyond the SHARDS wavelength range, we also make use of ancillary broad band GOODS-N data from \textit{HST}/ACS \citep{Giavalisco2004b,Riess2007}, \textit{HST}/WFC3 \citep{Grogin2011,Koekemoer2011} and \textit{Spitzer}/IRAC \citep{Fazio2004,PerezGonzalez2005,PerezGonzalez2008,Ashby2015}, as available in the Rainbow Cosmological Surveys Database\footnote{Operated by the Universidad Complutense de Madrid (UCM), partnered with the University of California Observatories at Santa Cruz (UCO/Lick, UCSC).\\ \url{http://rainbowx.fis.ucm.es/Rainbow_navigator_public/}.} \citep{Barro2011,Barro2011b,Barro2019}. The NIR data is particularly relevant when modelling these galaxies, since it provides more robust estimations of the ages and masses of any significant older SPs. Likewise, the \mbox{non-detection} in these NIR bands is typically linked to younger and/or less massive galaxies. The IRAC photometry, however, presents large Point Spread Function (PSF) sizes, which could lead to neighbour emission contamination. To correct this effect, the IRAC photometry available in Rainbow \citep{Barro2019} made use of the {\sc tfit} software \citep{Laidler2007}. This code takes accurate positions of the sources in the highest resolution band (\textit{HST}/F160W) and creates PSF-matched models of the objects in the lower-resolution bands, allowing the rejection of any flux contamination due to neighbour sources. For more details about the Rainbow photometry calculation we refer to \citet{Barro2019}.

\section{Methods}
\label{sec:Methods}
In order to shed light into the nature and evolution of LAEs and LBGs, we have used the Code Investigating GALaxy Emission \citep[{\sc cigale},][]{Noll2009, Boquien2019}. This python software builds stellar populations from synthetic models combined with various Star Formation Histories (SFHs). {\sc cigale} calculates the emission from gas ionised by massive stars, applying an attenuation law to both the ionised gas and the stars with a differential attenuation between young and old stars. The energy absorbed is re-emitted by the dust at mid/far infrared wavelengths. Combining all the input parameters given, {\sc cigale} creates a grid of models that are compared with the observed data, checking  their likelihood and selecting the best fit for each object. This \mbox{best-fitting} model is then used to derive the main physical parameters. For more details about {\sc cigale}, we refer to \citet{Noll2009} and \citet{Boquien2019}.

\subsection{The models}
\label{sec:Models}
\begin{figure}
	\includegraphics[width=\columnwidth]{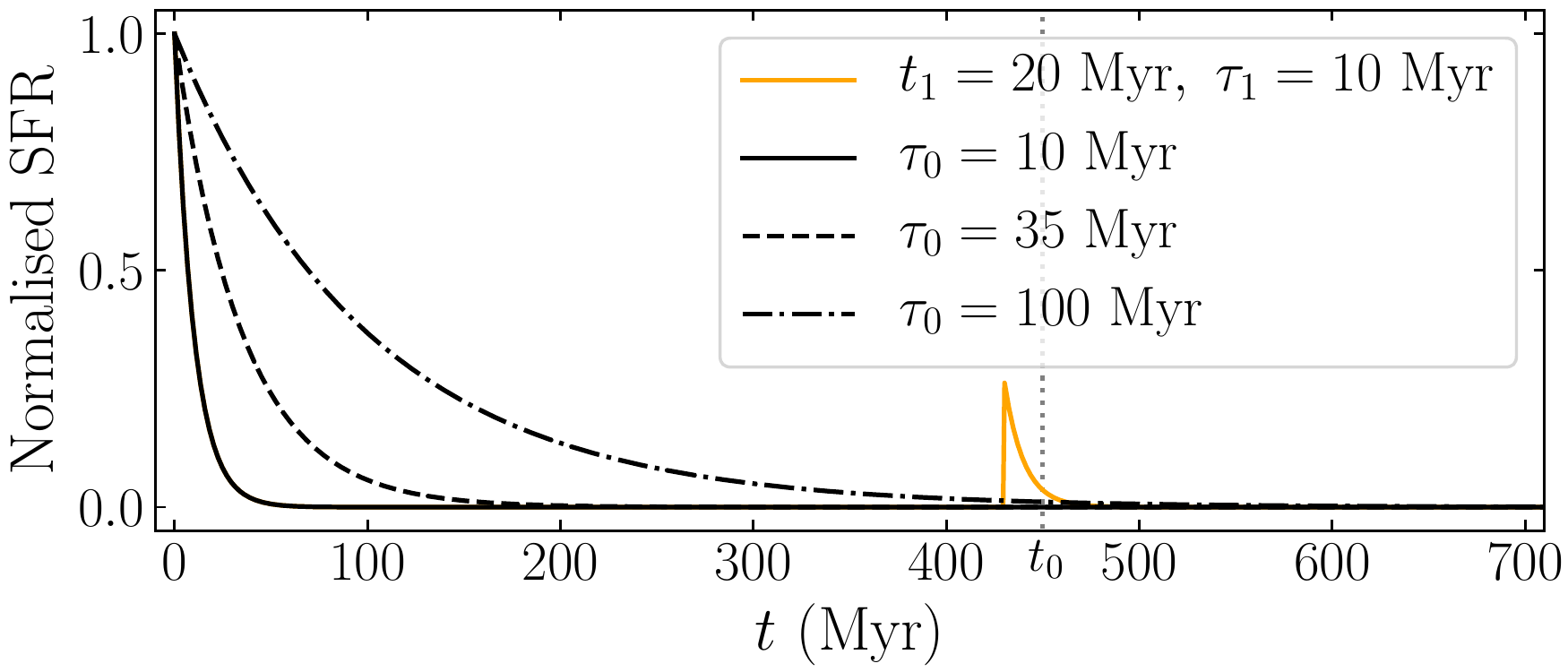}
   \caption{Double exponentially declining SFH for a main SP with different \mbox{$e$-folding} times (different line styles) presenting a second burst of star formation (solid orange line) for an arbitrary $f$ value. \mbox{$t=0$} corresponds to the formation of the galaxy, while $t_{0}$ (grey dotted vertical line) represents its current age and $t_{1}$ is the time elapsed since the beginning of the second burst of star formation to $t_{0}$.}
    \label{fig:SFH_plot}
\end{figure}

We use the commonly adopted exponentially declining SFH to model our SPs, as in, \textit{e.g.}, \citet{Papovich2001}, \citet{PerezGonzalez2003}, \citet{PerezGonzalez2008}, \citet{Serra2011}, \citet{RodriguezEspinosa2014} and \citet{Grazian2015} but see also \citet{Carnall2019} and \citet{Leja2019} for further discussions on SFHs. For this purpose, {\sc cigale} allows the use of a double exponential SFH consisting in a first decaying exponential corresponding to the long-term star formation responsible of the bulk of stellar mass, plus a second exponential that  models recent bursts of star formation. The combined SFHs can be expressed as follows:

\begin{equation}
\label{eq:exp_SFH}
 \mathrm{SFR}(t)\propto \begin{cases} \exp(-t/\tau_{0}) & \mbox{if } t<t_{0}-t_{1}\\ 
 \exp(-t/\tau_{0})+k\cdot \exp(-t/\tau_{1}) & \mbox{if } t\geq t_{0}-t_{1}, \end{cases}
\end{equation}\

\noindent where $\tau_{0}$ and $\tau_{1}$ are the \mbox{$e$-folding} times of the old and young exponential SPs, respectively, and $k$ is a constant indicating the relative strength of the young burst. The time $t_{1}$ is the age of the young population, while $t_{0}$ is so for the old one (see Fig.~\ref{fig:SFH_plot}). Furthermore, the fraction of stars formed in the young SP relative to the total stellar mass is given by the burst strength $f$, which can be expressed using discrete integrals, as {\sc cigale} accounts for the SFH with a period of \mbox{1 Myr}:

\begin{equation}
\label{eq:burst_str}
 f=\frac{k\sum_{t=t_{0}-t_{1}-1}^{t_{0}-1}\exp(-t/\tau_{1})}{\sum_{t=0}^{t_{0}-1}\exp(-t/\tau_{0})+k\sum_{t=t_{0}-t_{1}-1}^{t_{0}-1}\exp(-t/\tau_{1})}.
\end{equation}\

With this definition, $k$ can be written in the following way:

\begin{equation}
\label{eq:k_param}
 k=\frac{f}{1-f}\cdot\frac{\sum_{t=0}^{t_{0}-1}\exp(-t/\tau_{0})}{\sum_{t=t_{0}-t_{1}-1}^{t_{0}-1}\exp(-t/\tau_{1})},
\end{equation}\

\noindent which indeed leads to the classical case of a single exponential model when \mbox{$f=0$}.

Using this SFH, the models are computed with the \citet{Bruzual2003} stellar emission library, adding nebular templates based on \citet{Inoue2011}. A \citet{Salpeter1955} IMF is assumed as well as a \citet{Calzetti2000} dust extinction law. In order to avoid degeneracy and save computational time, we take the next approximations to constrain some of the many possible input physical parameters:

\begin{itemize}
\item Regarding \mbox{$e$-folding} times, a first test was made using a wide range of $\tau$ values of up to 1 Gyr, finding that a large majority of sources are better fitted with short \mbox{$e$-folding} times (see Fig.~\ref{fig:tau_dist}). In order to preserve the same nature of the SFH for the entire sample, a \mbox{$\tau\leq10$ Myr} constrain was adopted in the models, both for $\tau_{0}$ and $\tau_{1}$. Note that this short $\tau$ values are consistent with previous models of \mbox{high-$z$} galaxies as, \textit{e.g.}, \citet{RodriguezEspinosa2014} and \citet{HernanCaballero2017}, where fits prefer low values while allowing $\tau$ to vary. These short $\tau$ values correspond to SPs representing bursts of star formation.

\begin{figure}
	\includegraphics[width=\columnwidth]{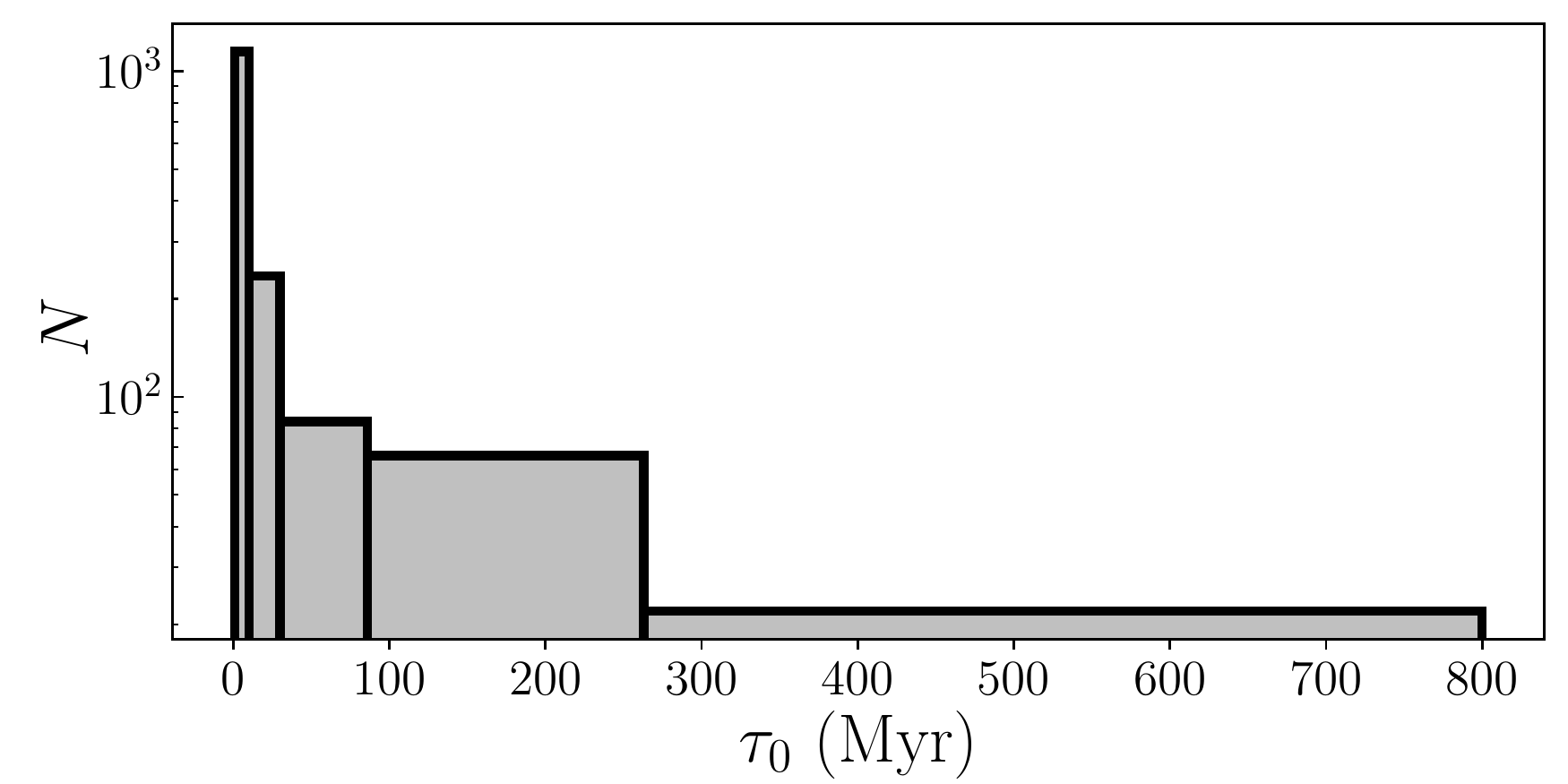}
   \caption{Distribution of the main SP \mbox{$e$-folding} times of the \mbox{best-fitting} models obtained leaving $\tau$ free up to 1 Gyr. Logarithmic bin widths are used for a better visualisation of the distribution at \mbox{$\tau\gtrsim10$ Myr}.}
    \label{fig:tau_dist}
\end{figure}

\item The Ly$\alpha$ escape fraction ($f_{\mathrm{esc}}$) is set to 0.15, consistent with previous calculations at \mbox{high-$z$} \citep{Robertson2010, Hayes2011, RodriguezEspinosa2014, Matthee2016, Sobral2017, Sobral2018b}. Very different escape fraction values are estimated in most recent works. \citet{Sobral2018b} show that the Ly$\alpha$ escape fraction for luminous \mbox{$z\sim2$-3} LAEs is very large (\mbox{$f_{\mathrm{esc}}\sim0.5$}). The Ly$\alpha$ escape fraction for common LAEs at \mbox{$z=2.23$} is also relatively high according to \citet{Sobral2017}, who measured a \mbox{$f_{\mathrm{esc}}\sim0.37$} by directly measuring H$\alpha$ and Ly$\alpha$ for these LAEs. On the other side, \citet{Matthee2016} also studied the escape fraction at \mbox{$z=2.23$} for more massive, star forming and dusty H$\alpha$ emitters, reporting much lower values (\mbox{$f_{\mathrm{esc}}\sim0.02-0.05$}). The cosmic average of the Ly$\alpha$ escape fraction is estimated around \mbox{$f_{\mathrm{esc}}\sim0.05-0.1$} \citep{Hayes2010, Sobral2017}. Note that the information available in our SEDs does not allow a robust estimation of $f_{\mathrm{esc}}$ for each galaxy and so leaving it as a free parameter would introduce a degeneracy with the age of the young SP. We instead adopt a value of 0.15, which is consistent with the estimation of $f_{\mathrm{esc}}$ through the mean Ly$\alpha$ EW$_{0}$ of the sample using the empirical estimator from \citet{Sobral2019}.

\item For the dust correction, the colour excess is assumed to be relatively low, up to \mbox{$\mathrm{E}(B-V)=0.12$}. This agrees with the values of this parameter calculated using the mean $\beta$ slopes, and the $A_{\mathrm{uv}}$ obtained in \citet{ArrabalHaro2018} in the $z$ range of study.

\item For the Single Stellar Population (SSP) fits (burst strength of \mbox{$f=0$}), the age is let free within a logarithmic range from 3 to \mbox{1500 Myr}. On the other hand, the Double Stellar Population (DSP) models have the age of their burst limited to a maximum of \mbox{50 Myr}, while the age of the underlying population can vary in a logarithmic range from that age to \mbox{1500 Myr}. Several tests with different age ranges were made to constrain these values. Those carried out with maximum ages below \mbox{1500 Myr} showed a peak and an abrupt cut at the maximum allowed age, especially at the lowest redshifts. At the same time, no galaxies presented ages above \mbox{1500 Myr} when the maximum age limit was further extended, which was expected since this value is close to the age of the Universe at our lowest redshift. None of the old SPs from the DSP models were neither younger than \mbox{50 Myr} when this lower limit was extended.

\item The relative strength of the young starburst in the DSP models is limited between a minimum burst strength of \mbox{$f=0.005$} and a maximum of \mbox{$f=0.5$}, which means that the stellar mass of the young population should represent at least a 0.5\% of the total stellar mass of the galaxy in the DSP best fits.

\item We allow the metallicity to vary from \mbox{$Z=10^{-4}$} up to \mbox{$Z=Z_{\odot}$}.
\end{itemize}

Finally, we make use of a {\sc cigale} feature that allows us to specify a prior for the integrated Ly$\alpha$ line flux, previously measured in \citet{ArrabalHaro2018}. In this way, the Ly$\alpha$ line is weighted more heavily than other data-points. Thereby, we make sure that the Ly$\alpha$ line is well modelled in those galaxies presenting it, avoiding low $\chi^{2}$ solutions where the SED is well fitted except for the filter detecting the line, which otherwise would be selected as \mbox{best-fitting} solutions but that actually do not represent the Ly$\alpha$ emission well. In case of the 1,030 \mbox{no-Ly$\alpha$} LBGs, to avoid imposing too strong constrains in the \mbox{non-detected} Ly$\alpha$ emission, we provide them with a common negligible input Ly$\alpha$ flux value, while assigning large enough errors to reach the integrated flux of the faintest Ly$\alpha$ line flux measured in the LAEs sample with the SHARDS filters in \citet{ArrabalHaro2018}, \textit{i.e.}, \mbox{$F_{\mathrm{min}}(\mathrm{Ly}\alpha)\simeq1.3^{+2.0}_{-1.3}\times 10^{-19}\ \mathrm{erg\ s^{-1}\ cm^{-2}}$}.

\begin{figure}
	\includegraphics[width=0.49\columnwidth]{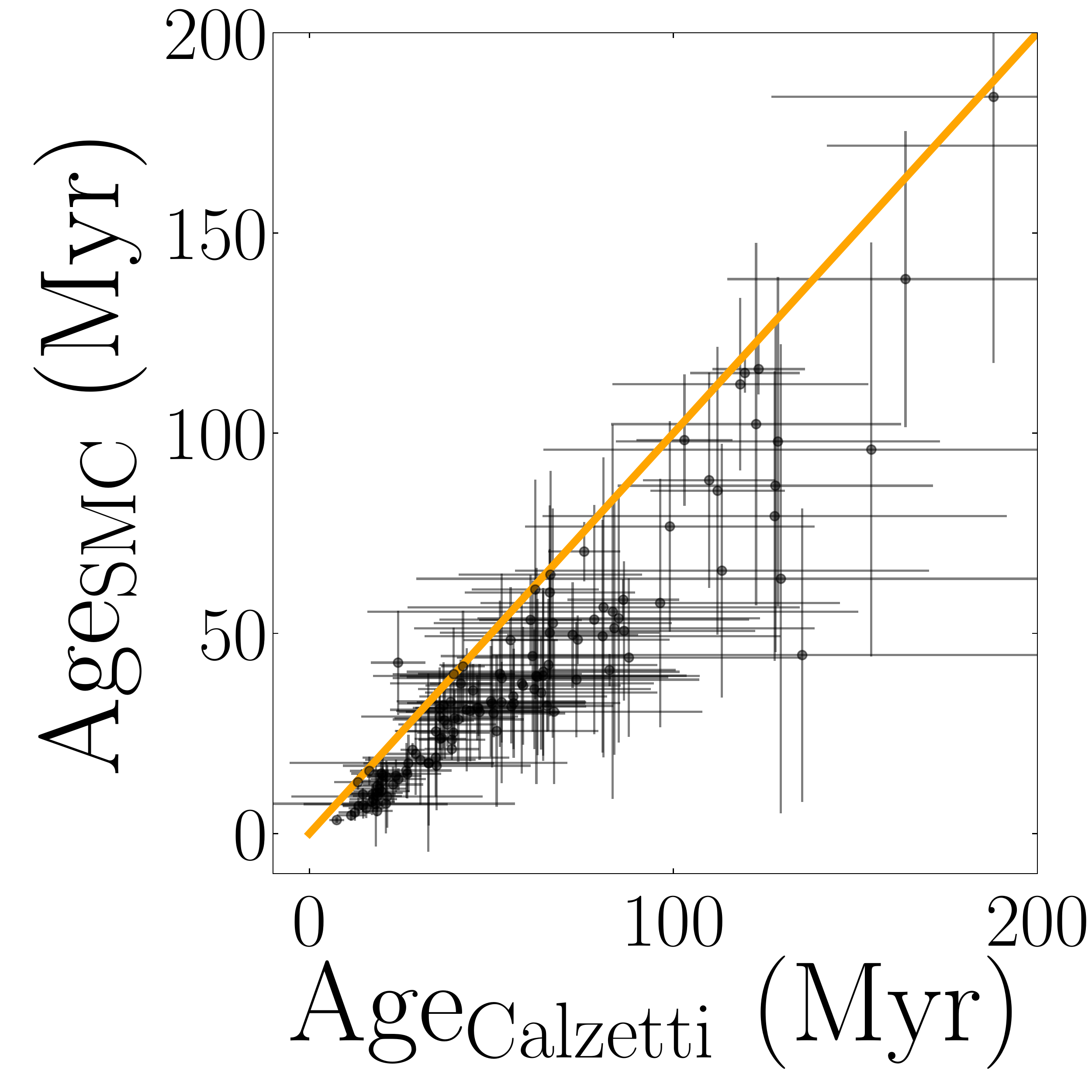}
	\includegraphics[width=0.49\columnwidth]{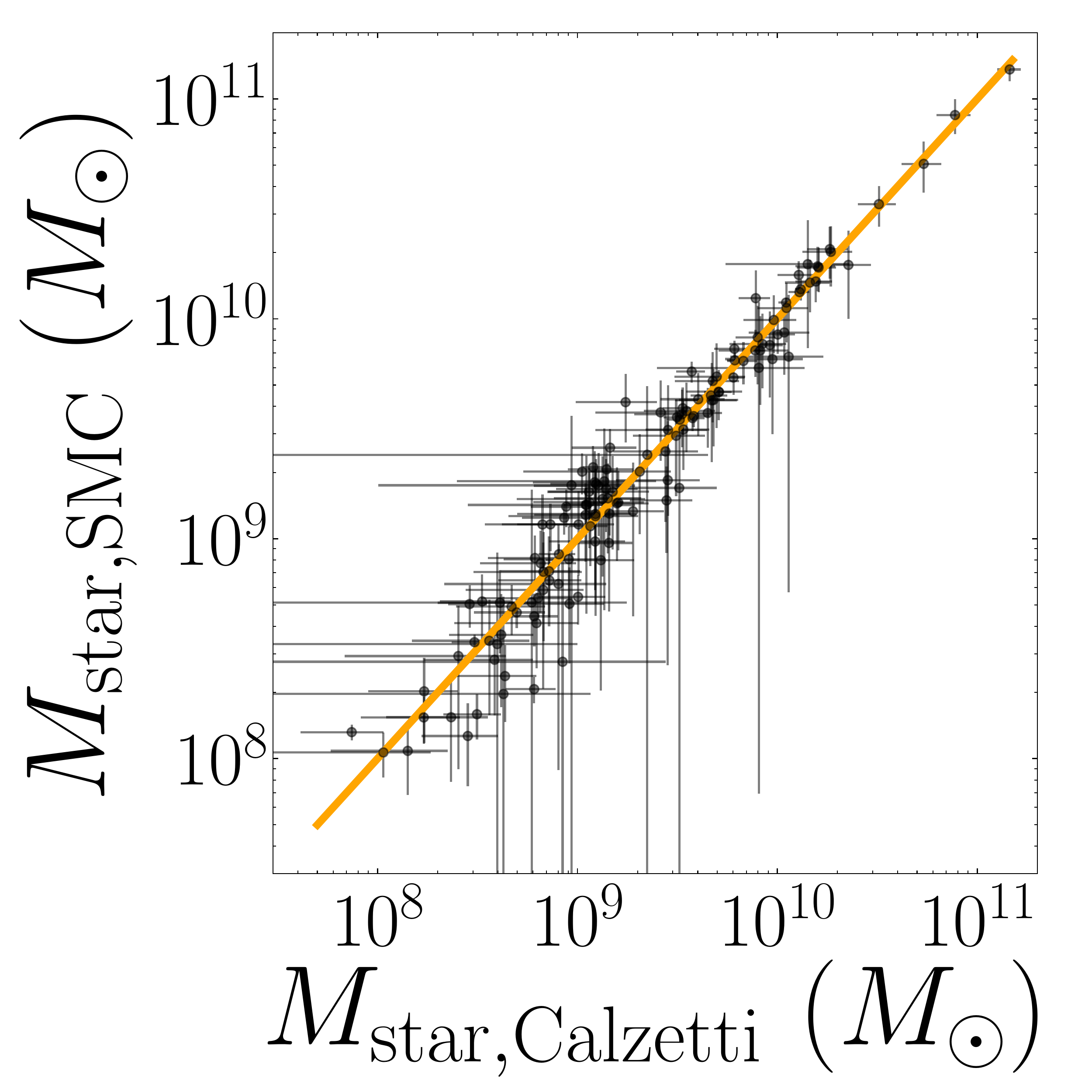}
  \caption{Age and stellar mass differences between a Calzetti and a \mbox{SMC-like} dust extinction law for a random subsample. The one-to-one line is shown in orange. Only ages below \mbox{200 Myr} are shown for clarity.}
    \label{fig:dust_diff}
\end{figure}

Additionally, in order to explore how the choice of a different dust extinction law affects the main physical parameters derived with the models, we carried out some tests with a randomly selected subsample of 132 sources (preserving the proportion of galaxies from each observational subclass). This subsample was fitted using single SP SFHs with the exact same parameters described in this section but varying from a \citet{Calzetti2000} dust law to a \mbox{SMC-like} dust law. The differences obtained in ages and stellar masses are shown in Fig.~\ref{fig:dust_diff}. It can be appreciated that ages estimated with the \citet{Calzetti2000} extinction law are systematically older. Nonetheless, the age difference is not very significant in terms of the associated error bars of the age estimations. 

\begin{figure}
	\includegraphics[width=0.49\columnwidth]{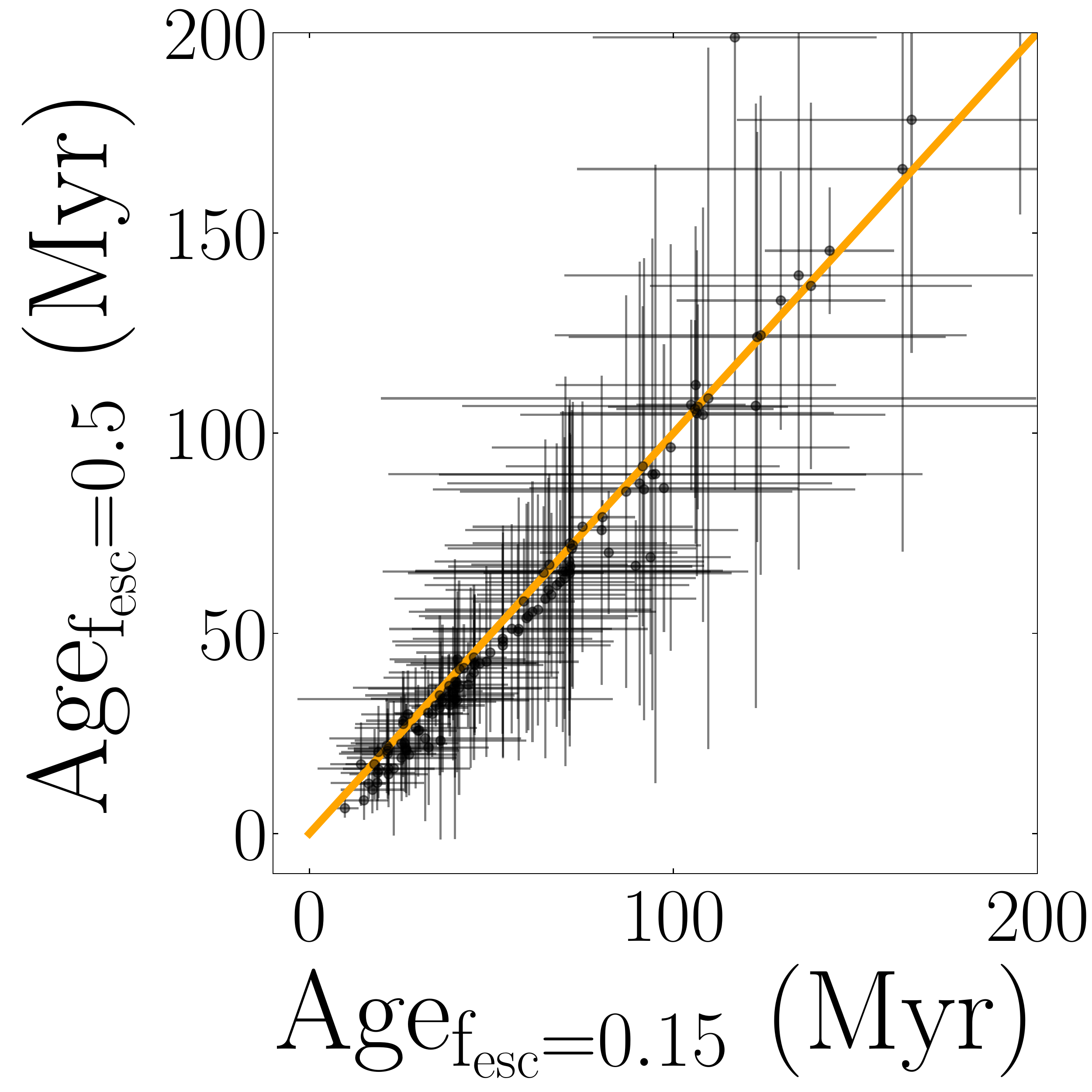}
	\includegraphics[width=0.49\columnwidth]{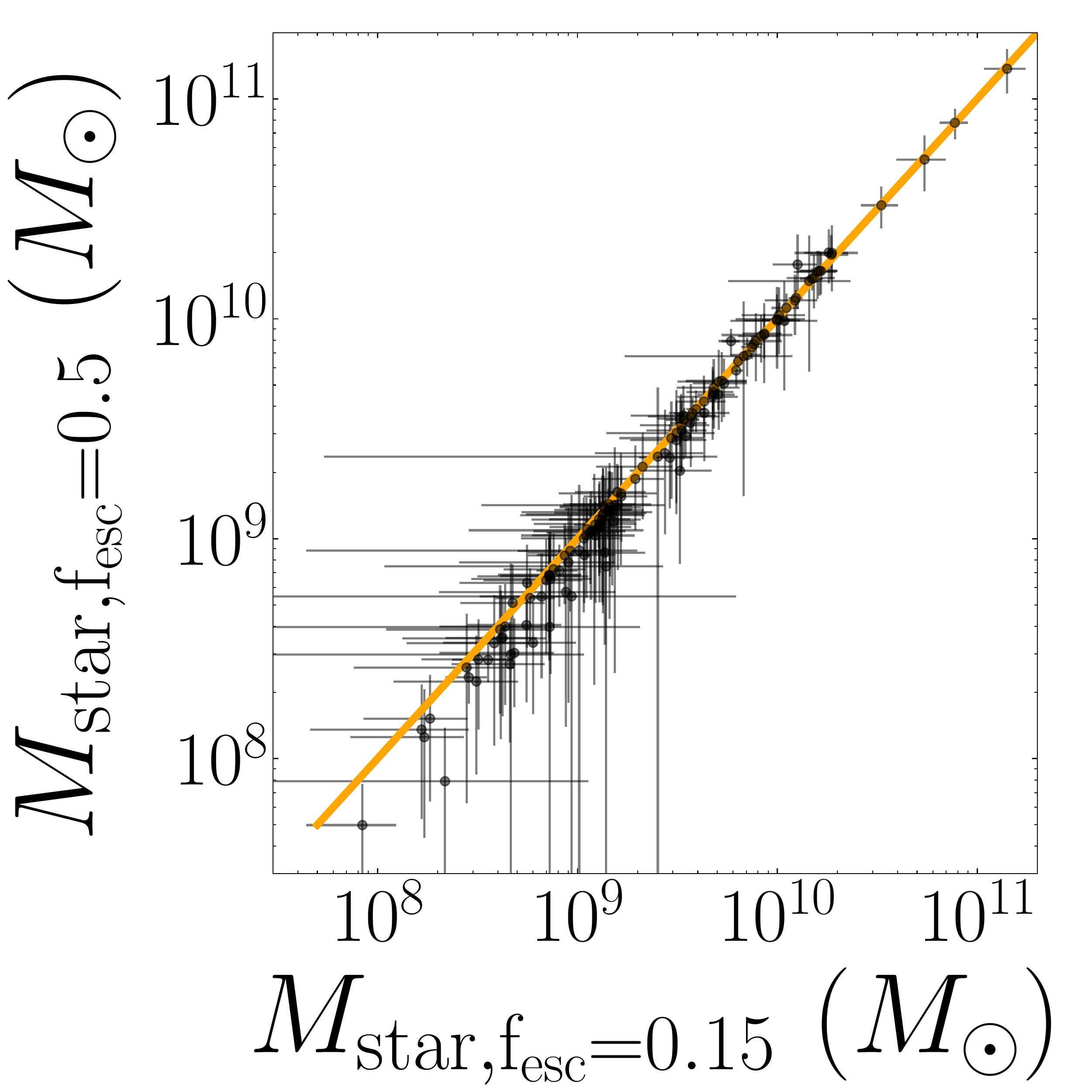}
  \caption{Age and stellar mass differences between \mbox{$f_{\mathrm{esc}}=0.15$} and \mbox{$f_{\mathrm{esc}}=0.5$} for a random subsample. The one-to-one line is shown in orange. Only ages below \mbox{200 Myr} are shown for clarity.}
    \label{fig:f_esc_diff}
\end{figure}

The same subsample of 132 objects was also fitted using a larger \mbox{$f_{\mathrm{esc}}=0.5$}. Fig.~\ref{fig:f_esc_diff} shows the age and stellar mass differences found. It can be noted that the derived ages are slightly older when using \mbox{$f_{\mathrm{esc}}=0.15$}, especially for the youngest objects, although the age difference is again not relevant if the typical uncertainties of this parameter are taken into account. Moreover, the youngest objects are this time slightly less massive as $f_{\mathrm{esc}}$ increases. This is consistent, since given an observed Ly$\alpha$ flux, it would correspond to a fainter Ly$\alpha$ intrinsic luminosity the higher the $f_{\mathrm{esc}}$ is. The same test was carried out for the entire pure LAEs sample, more prone to have higher $f_{\mathrm{esc}}$ since they present higher Ly$\alpha$ EWs \citep{Sobral2019}. We find the same behaviour. In any case, the stellar masses obtained using different extinction laws and escape fractions are very similar and so no specially significant changes will take place in this regard, even though it is worth noticing that the ages obtained would be shorter if we used the \mbox{SMC-like} dust law and a larger $f_{\mathrm{esc}}$ for the analysis of the entire sample.
 
\begin{figure}
	\includegraphics[width=\columnwidth]{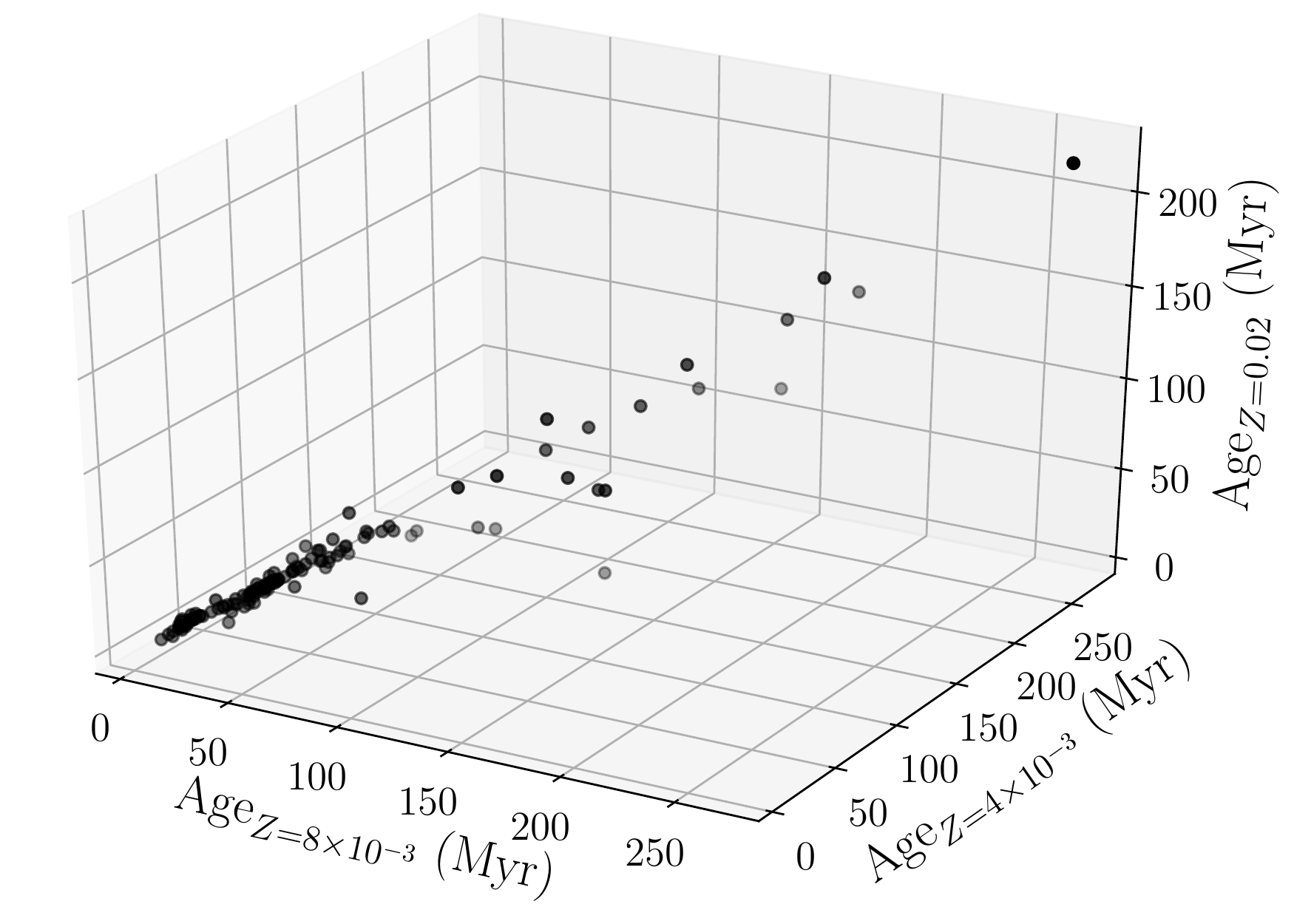}
  \caption{Age comparison of a random subsample for three different fixed metallicities, using three typical values of the general sample. Error bars have been omitted for clarity, but their magnitude is comparable with that in the left panels of Figs.~\ref{fig:dust_diff} and \ref{fig:f_esc_diff}.}
    \label{fig:metal_diff}
\end{figure}

Furthermore, herein we will use an interpretation of the ages in relative terms. This is because of their large uncertainties, especially for the oldest galaxies. Since the existing degeneracy between age and metallicity could be relevant even when relatively comparing the derived ages, another test was made with the 132 random subsample, making several fits fixing a unique metallicity value each time (see Fig.~\ref{fig:metal_diff}). The ages derived show that younger galaxies keep being younger independently of the metallicity. Note that the age-metallicity degeneracy can still be relevant for objects in the edges of the metallicity range employed for the models, although the age differences are within the typically large age errors.

\subsection{Single and double stellar population considerations}
\label{sec:one_or_two_SP}
Since we want to determine whether a second SP is needed to model our SHARDS \mbox{high-$z$} galaxies, both SSP and DSP models are run separately, and their best solutions compared.

To discriminate between the two approaches, we make use of the Bayesian Information Criterion \citep[BIC,][]{Schwarz1978} as explained in \citet{Liddle2007} and applied in, \textit{e.g.}, \citet{MendezAbreu2018}:

\begin{equation}
\label{eq:BIC}
 \mathrm{BIC}=\chi^{2}+q\ln(m),
\end{equation}\

\noindent where $q$ is the number of free parameters of the model used and $m$, the number of independent data points available. In our particular case, SSP models have five free parameters: age, stellar mass, \mbox{$e$-folding} time, colour excess and metallicity. However, $\mathrm{E}(B-V)$ and $Z$ are the same for the young and the old SPs in DSP models, and so those have eight free parameters.

The advantage of using this Bayesian indicator over the $\chi^2$ when comparing results from different models is that the BIC penalises the addition of extra free parameters in a stronger way than the normal or even the reduced $\chi^2$. Defining the BIC difference between SSP and DSP models as $\Delta\mathrm{BIC}\equiv\mathrm{BIC_{1SP}}-\mathrm{BIC_{2SP}}$, there is a $\Delta\mathrm{BIC}$ threshold from which higher $\Delta\mathrm{BIC}$ values correspond to scenarios where the additional free parameters (in this case, an extra SP) are needed to properly model the galaxy.

\begin{figure}
	\includegraphics[width=\columnwidth]{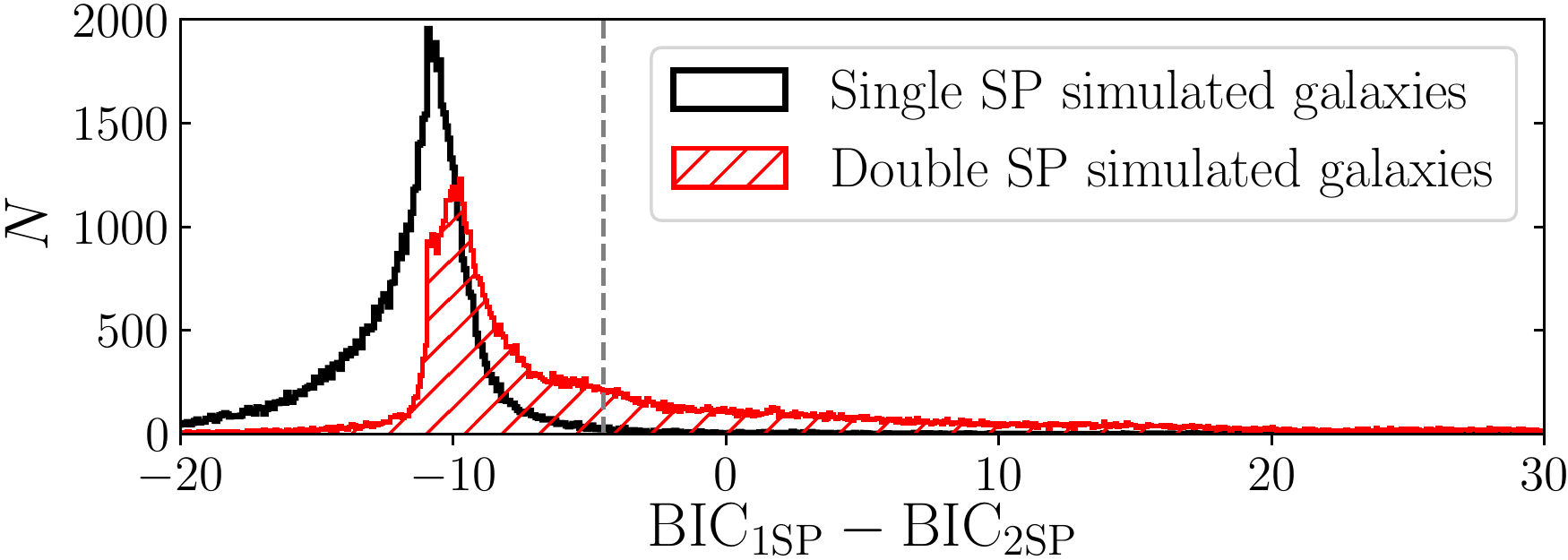}
   \caption{$\Delta\mathrm{BIC}$ distribution of the mock galaxies. The vertical dashed line marks the threshold \mbox{$\Delta\mathrm{BIC}=-4.50$} beyond which a source requires a DSP model. There is a majority of galaxies for which the SSP and DSP best solutions have similar $\chi^{2}$, resulting in \mbox{$\Delta\mathrm{BIC}\simeq-10$} for \mbox{$q_{1}=5$}, \mbox{$q_{2}=8$} at a typical number of independent SED points of \mbox{$m\sim30$} (see Eq.~\ref{eq:BIC}). For objects with \mbox{$\Delta\mathrm{BIC}<-4.50$}, SSP models are preferred over double ones.}
    \label{fig:mock_d_BIC}
\end{figure}

To calibrate our $\Delta\mathrm{BIC}$ and obtain the threshold value, a set of theoretical models created with one and two SPs are all fitted with {\sc cigale} using SSP and DSP models in a separated way in order to compare their $\Delta\mathrm{BIC}$ distributions, following the method used in \citet{MendezAbreu2018}. In particular, we take the \mbox{best-fitting} SSP model and the \mbox{best-fitting} DSP model of each one of the actual galaxies, convolve them through our photometry filters and use Monte Carlo simulations to create 50 new mock samples by perturbing the convolved photometry with Gaussian noise in accordance with the photometric error of each point. In this way, we obtain \mbox{$\sim$155,800} mock SEDs whose origin is known (half of them are product of SSP models and the other half come from DSP models). Moreover, these mock SEDs provide a good representation of our observed \mbox{high-$z$} galaxy sample as they are perturbations of the convolution of the best models fitted to the actual sample. They are then modelled both with single and double SP in order to compare the resulting $\Delta\mathrm{BIC}$ distribution. The obtained histogram of $\Delta\mathrm{BIC}$ values for the mock galaxies is shown in Fig.~\ref{fig:mock_d_BIC}. The limit from which 95.45\% (2$\sigma$ significance) of the models come from DSP simulations is given by \mbox{$\Delta\mathrm{BIC}=-4.50$}. Those cases for which \mbox{$\Delta\mathrm{BIC}>-4.50$} can therefore be selected as our \textit{bona fide} DSP galaxies. Note that sources with \mbox{$\Delta\mathrm{BIC}<-4.50$} might still be DSP galaxies. However, we cannot precisely discern the best way of modelling each one of those individual sources in terms of their $\Delta\mathrm{BIC}$, and so the simplest model is favoured over the more complex one. That is, the best SSP fit is taken as the best model for galaxies with \mbox{$\Delta\mathrm{BIC}<-4.50$}, while for those with \mbox{$\Delta\mathrm{BIC}>-4.50$} the best DSP model is taken. Notice also that the $\Delta\mathrm{BIC}$ value at which the purity of DSP objects reaches the 2$\sigma$ level could vary depending on the proportion of generated SSP and DSP models. In this regard, using a set of Monte Carlo perturbations of both the SSP and DSP \mbox{best-fitting} solution of each original galaxy of the sample is especially relevant to estimate the threshold $\Delta\mathrm{BIC}$ value for this particular sample.

\section{Results}
\label{sec:Results}
With the caveats mentioned in Sec.~\ref{sec:Methods}, we obtained good {\sc cigale} solutions for all the sources in the sample but three very faint ones whose SEDs lack detection in many filters and present large uncertainties in the measurements. In this section we present the results of the model fitting, highlighting the most significant differences found between the three observational classes in our sample. Even though many physical parameters are computed by {\sc cigale} during the model fitting, we will focus here only in the ages and stellar masses of the sample. We note that there is a well known degeneracy between age, dust and metallicity, and therefore age values should be taken with care in an absolute sense.

In Table~\ref{tab:results} we present the IDs, coordinates, ages and stellar masses (split into young and old SP when we select a DSP synthetic spectrum) derived for our 1,555 well modelled galaxies. The median age and $M_{\mathrm{star}}$ of each subgroup are shown in Table~\ref{tab:median_results}.

\begin{table*}
    \centering
    \caption{Main physical parameters derived from the best models fitted to each source: name of the object using the SHARDS identification, right ascension and declination, photometric redshift, stellar mass, age and \mbox{$e$-folding} time of the main SP and same parameters of the second younger population (when needed). The total $M_{\mathrm{star}}$ is the sum of $M_{\mathrm{star,m}}$ and $M_{\mathrm{star,b}}$. The last three column are left empty when the best solution does not need any additional SP. A full version of this table is available in the on-line version.}
    \label{tab:results}
    \begin{tabular}{cccccccccc}
    \hline
    Object Name & R.A. & Dec. & $z$ & $M_{\mathrm{star,m}}$ & $\mathrm{Age_{m}}$ & $\tau_{0}$ & $M_{\mathrm{star,b}}$ & $\mathrm{Age_{b}}$ & $\tau_{1}$\\
     & (J2000) & (J2000) & & ($10^{9}\ M_{\odot}$) & (Myr) & (Myr) & ($10^{9}\ M_{\odot}$) & (Myr) & (Myr)\\    
\hline
SHARDS20010117 & 12:35:48.1 & 62:12:02.4 & $4.28\pm0.06$ & $1.88\pm1.24$ & $77\pm44$ & $3.4\pm2.7$ & - & - & -\\
SHARDS20007539 & 12:35:48.1 & 62:12:03.8 & $5.38\pm0.07$ & $2.34\pm1.04$ & $24\pm12$ & $4.24\pm2.35$ & - & - & -\\
SHARDS20012481 & 12:35:50.9 & 62:11:58.5 & $5.69\pm0.06$ & $2.74\pm1.51$ & $63\pm31$ & $2.91\pm2.32$ & - & - & -\\
SHARDS20005927 & 12:35:51.5 & 62:12:16.5 & $3.22\pm0.07$ & $2.8\pm0.5$ & $47\pm10$ & $2.36\pm1.51$ & - & - & -\\
SHARDS20005405 & 12:35:51.6 & 62:12:12.7 & $4.03\pm0.07$ & $6.5\pm1.31$ & $74\pm21$ & $2.92\pm2.2$ & - & - & -\\
SHARDS20008074 & 12:35:52.2 & 62:11:20.8 & $5.53\pm0.07$ & $0.60\pm0.32$ & $20\pm9$ & $1.9\pm1.56$ & - & - & -\\
SHARDS20008444 & 12:35:53.2 & 62:10:32.9 & $4.01\pm0.07$ & $6.3\pm5.0$ & $950\pm500$ & $5.0\pm3.2$ & $0.89\pm1.30$ & $35\pm12$ & $2.02\pm1.5$\\
SHARDS20010810 & 12:35:53.4 & 62:10:23.3 & $5.12\pm0.06$ & $1.4\pm0.9$ & $72\pm44$ & $3.3\pm2.7$ & - & - & -\\
SHARDS20005669 & 12:35:54.1 & 62:10:32.9 & $3.36\pm0.07$ & $11.0\pm7.0$ & $357\pm244$ & $5.2\pm3.3$ & $0.46\pm0.94$ & $24\pm14$ & $3.0\pm2.9$\\
SHARDS20011405 & 12:35:54.3 & 62:10:18.8 & $5.37\pm0.07$ & $3.69\pm1.44$ & $80\pm38$ & $3.6\pm2.7$ & - & - & -\\
SHARDS20006420 & 12:35:54.4 & 62:10:33.8 & $3.88\pm0.06$ & $7.06\pm2.49$ & $110\pm50$ & $4.3\pm3.0$ & - & - & -\\
SHARDS20006258 & 12:35:54.5 & 62:12:14.6 & $3.48\pm0.06$ & $1.1\pm0.6$ & $17\pm11$ & $2.04\pm1.78$ & - & - & -\\
SHARDS20013727 & 12:35:55.0 & 62:12:04.8 & $5.96\pm0.07$ & $20.0\pm10.0$ & $294\pm190$ & $5.2\pm3.3$ & $1.12\pm2.08$ & $32\pm15$ & $4.8\pm3.3$\\
SHARDS20009009 & 12:35:55.2 & 62:11:25.4 & $3.89\pm0.06$ & $1.7\pm0.6$ & $66\pm28$ & $3.2\pm2.48$ & - & - & -\\
SHARDS20006827 & 12:35:55.7 & 62:10:19.0 & $4.28\pm0.06$ & $1.9\pm0.8$ & $38\pm16$ & $1.83\pm1.21$ & - & - & -\\
... & ... & ... & ... & ... & ... & ... & ... & ... & ...\\
  \hline
    \end{tabular}
\end{table*}

\begin{table}
    \centering
     \caption{Median age and stellar mass for each group and subgroup in the sample. Ages shown correspond to the age of the old SP. Error bars correspond to the standard deviation of the median.}
    \label{tab:median_results}
    \begin{tabular}{lcc}
    \hline
        Type   &   Age (Myr)  &   $M_{\mathrm{star}}$ ($10^{9}\ M_{\odot}$)\\
\hline
  No-Ly$\alpha$ LBGs     & $71\pm12$     & $3.5\pm1.1$\\
  LAE-LBGs        & $40\pm27$    & $2.3\pm1.7$\\
  Pure LAEs        & $26^{+41}_{-25}$    & $0.56^{+1.20}_{-0.55}$\\
  \hline
    \end{tabular}
\end{table}

Additionally, as previously discussed in Sec.~\ref{sec:Data}, the availability of IRAC detection in the SEDs is especially relevant to discern whether we need to add a second SP or not, as well as to reliably constrain the main physical parameters derived. In Table~\ref{tab:SP_needed} we give an overview of the proportion of objects within each subfamily with detection in IRAC as well as the fraction of sources requiring the addition of a second SP. Indeed, \mbox{$\sim$98\%} of the DSP galaxies are detected in IRAC. Moreover, these sources present median age and $M_{\mathrm{star}}$ relative errors of 40\% and 33\%, respectively, while these relative errors increase to 53\% and 58\% for the age and $M_{\mathrm{star}}$ of the \mbox{IRAC-undetected} objects.

\begin{table}
    \centering
     \caption{Number and fraction of objects from each family detected in IRAC and number of sources requiring two SPs to model their SEDs according to the $\Delta\mathrm{BIC}$ criterion.}
    \label{tab:SP_needed}
    \begin{tabular}{lccc}
    \hline
        Type  &   $N_{\mathrm{total}}$ & $N_{\mathrm{IRAC}}$  &   $N_{\mathrm{DSP}}$\\
\hline
  No-Ly$\alpha$ LBGs  & 1030    & 882 (86\%)  & 164 (15.9\%)\\
  LAE-LBGs    & 404    & 347 (86\%)    & 135 (33.4\%)\\
  Pure LAEs     & 124    & 71 (57\%)    & 27 (21.8\%)\\
  % Complete sample  & 326 (21.0\%)\\         
  \hline
    \end{tabular}
\end{table}

\subsection{Stellar populations required}
\begin{figure}
	\includegraphics[width=\columnwidth]{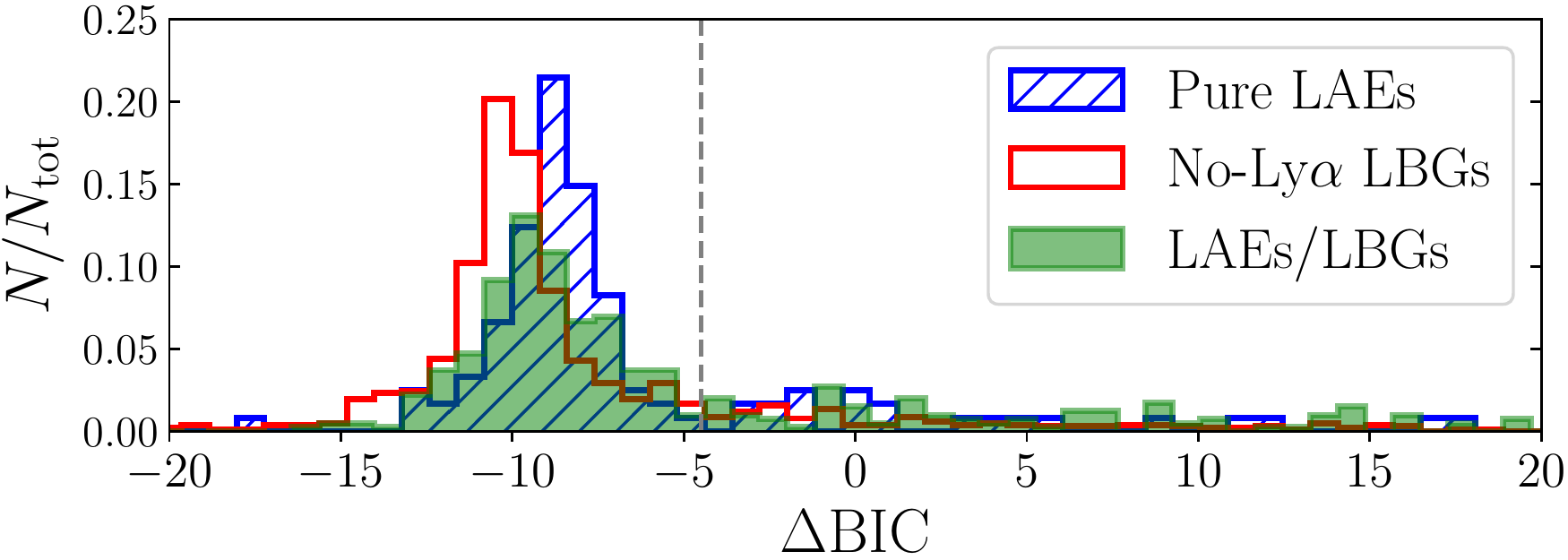}
   \caption{$\Delta\mathrm{BIC}$ distribution of pure LAEs (blue diagonals), LBGs with no Ly$\alpha$ emission line (red unfilled) and \mbox{LAE-LBGs} (solid green) weighted by the amount of objects within each class. The \mbox{$\Delta\mathrm{BIC}=-4.50$} value from which two SPs are needed is depicted by a vertical dashed line. It can be appreciated that the majority of objects (79.0\%) are well modelled using a SSP model. Thus, pure LAEs and \mbox{no-Ly$\alpha$} LBGs are proportionally less represented in the DSP models $\Delta\mathrm{BIC}$ range than \mbox{LAE-LBGs}, whose distribution is wider and more extended towards high $\Delta\mathrm{BIC}$ values.}
    \label{fig:d_BIC_families}
\end{figure}

We apply the calibrated $\Delta\mathrm{BIC}$ criterion explained in Sec.~\ref{sec:one_or_two_SP} to get the \mbox{best-fitting} model for each individual galaxy of the sample with either one or two SPs, finding that in most cases (79.0\%), these \mbox{high-$z$} galaxies do not require the addition of an extra SP to model their SEDs. The frequency distribution of $\Delta\mathrm{BIC}$ for the three families is shown in Fig.~\ref{fig:d_BIC_families}. Even though the three groups clearly present their peaks within the SSP $\Delta\mathrm{BIC}$ range, it is worth noticing that the \mbox{LAE-LBGs} family is wider and the one that more likely tends towards higher $\Delta\mathrm{BIC}$ values. Specifically, only 15.9\% and 21.8\% of the \mbox{no-Ly$\alpha$} LBGs and pure LAEs need of a second SP in their fits, respectively, while this occurs for 33.4\% of the \mbox{LAE-LBGs} (see Table~\ref{tab:SP_needed}). Further discussion on the reasons of this behaviour will be given in next subsections and Sec.~\ref{sec:Discussion}.

\subsection{Age differences}
\begin{figure*}
	\includegraphics[width=\textwidth]{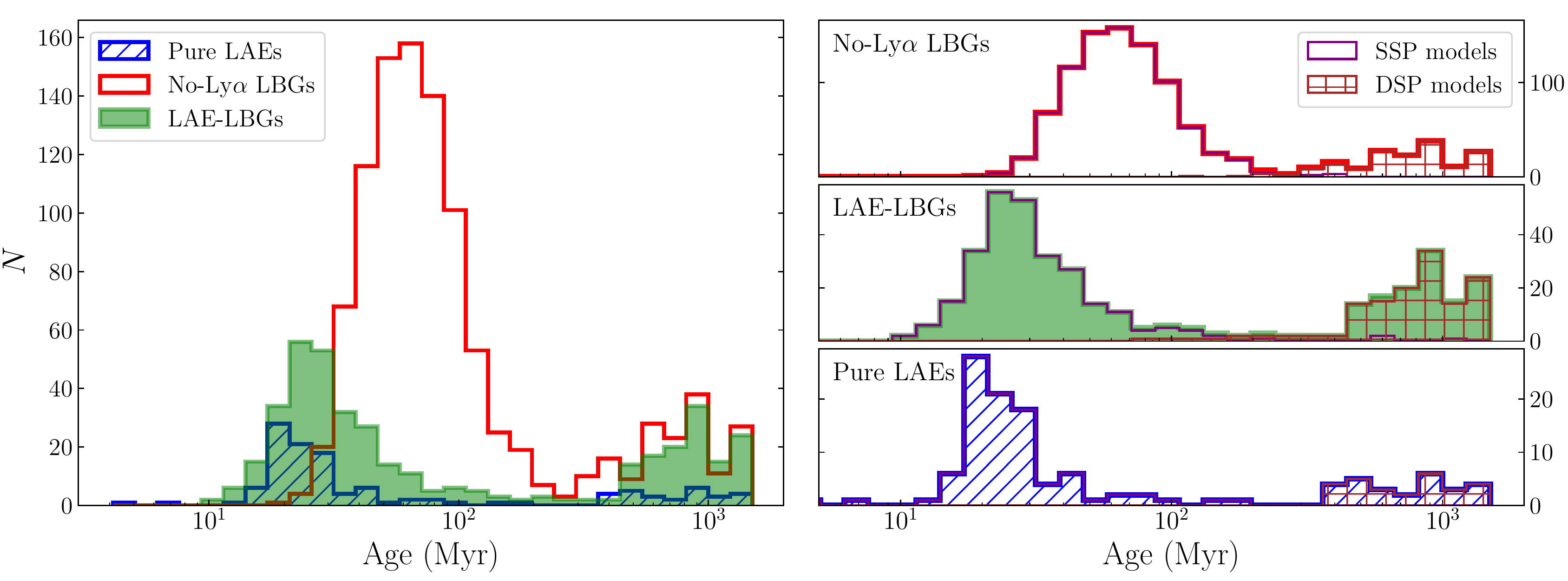}
   \caption{\textit{Left panel}: Main SP age distribution for the pure LAEs (blue diagonals), \mbox{LAE-LBGs} (solid green) and no Ly$\alpha$ line LBGs (unfilled red). Notice that most pure LAEs show low ages (71\% below \mbox{50 Myr}), while the \mbox{no-Ly$\alpha$} LBGs are typically older, peaking at \mbox{$\sim71$ Myr}. \textit{Right panels}: Split contribution of SSP and DSP models to the age distribution of each family. The observed dichotomy is due to the use of \mbox{burst-like} SFHs joined to the limitations of the SEDs to constrain the old SP in DSP models.}
    \label{fig:age_dist}
\end{figure*}

Looking at the age distribution in Fig.~\ref{fig:age_dist}, a large difference can be appreciated between pure LAEs and \mbox{no-Ly$\alpha$} LBGs, where the first ones are much younger, with a median age of \mbox{$26^{+41}_{-25}$ Myr}, while that of the \mbox{no-Ly$\alpha$} LBGs is \mbox{$71\pm12$ Myr}. It can also be appreciated a clear dichotomy in the age distribution led by SSP and DSP models. This dichotomy is indeed product of the use of \mbox{burst-like} SFHs joined to the large uncertainties associated to the physical parameters of the old SP in DSP models. In the cases where two SPs are needed to reproduce the SED, the Ly alpha line and UV continuum are well fitted by the young and well defined SP and so the old SP can adopt a large variety of ages in order to fit the continuum points at longer wavelengths. When this happens, very large ages typically lead to the best $\chi^2$ of the global fit, producing a not representative gap of sources at intermediate ages. The discussion in this work is however focused on the relative differences between SSP and DSP sources, but the exact values of the age for the DSP models are too uncertain to consider them as accurate absolute calculations, but only as an estimation of the order of magnitude.

Even though the age dichotomy is found for the three families, special attention is put onto the \mbox{LAE-LBGs} as their SEDs present emission features better constraining the young and old SP and therefore the need of a DSP model: Ly$\alpha$ emission line (not present in \mbox{no-Ly$\alpha$} LBGs) and bright continuum measure, especially relevant at long wavelengths (not present in pure LAEs). Moreover, this subgroup is the one that more frequently needs the use of DSP models. The two peaks of the \mbox{LAE-LBGs} age distribution can be associated to the inherent nature of the models fitting those objects, split into: 1) SSP \mbox{LAE-LBGs}, corresponding to very young galaxies (median \mbox{age $\sim27$ Myr}) with large enough $M_{\mathrm{star}}$ to show a prominent UV continuum detectable in SHARDS, and 2) DSP \mbox{LAE-LBGs}, representing a more evolved galaxy with an older underlying massive SP suffering a recent star-forming episode, thus the additional young SP.

To shed light on whether the SSP age difference between \mbox{no-Ly$\alpha$} LBGs and \mbox{LAE-LBGs} is only driven by the detection of the Ly$\alpha$ line or not, SSP \mbox{LAE-LBGs} are fitted a second time omitting the Ly$\alpha$ line contribution from their SEDs by replacing the flux in the filter sampling the Ly$\alpha$ line with an estimation of the continuum emission from the adjacent SHARDS filters or the broad band \textit{HST}/ACS photometry when needed. The new Ly$\alpha$-removed photometry is then refitted with and without the integrated Ly$\alpha$ flux prior used for the \mbox{no-Ly$\alpha$} LBGs (see Sec.~\ref{sec:Models}). In both cases, the median age obtained for the SSP \mbox{LAE-LBGs} in this second run is only \mbox{$\sim6$-8 Myr} younger than that of the SSP \mbox{no-Ly$\alpha$} LBGs, being not different within their errors. This highlights that the youthfulness of the SSP \mbox{LAE-LBGs} in the models comes from the Ly$\alpha$ emission line, also implying that some \mbox{no-Ly$\alpha$} LBGs for which the intrinsic Ly$\alpha$ line remains undetected because of scattering or dust extinction could be equally young as well, adding an extra difficulty in the characterisation of this last subgroup.

\subsection{Stellar mass differences}
\begin{figure*}
	\includegraphics[width=\textwidth]{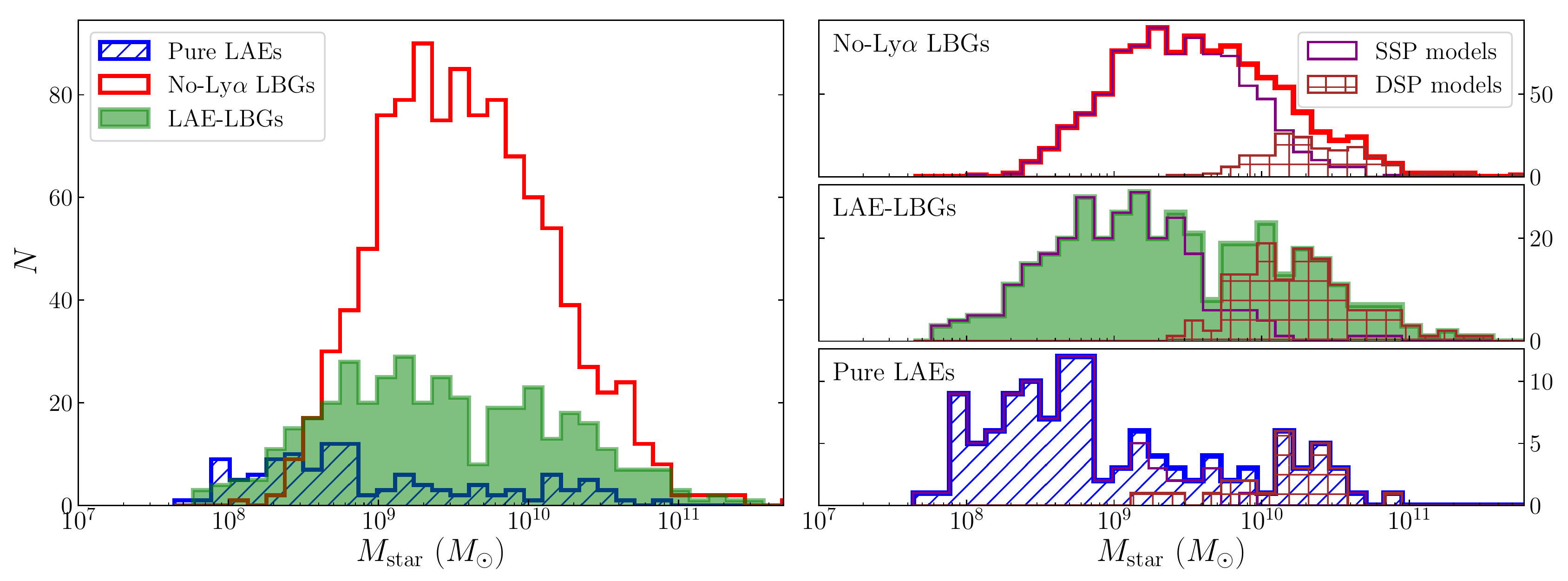}
   \caption{\textit{Left panel}: Stellar mass distribution of pure LAEs (blue diagonals), \mbox{LAE-LBGs} (solid green) and \mbox{no-Ly$\alpha$} LBGs (unfilled red). Notice that both LBGs families show substantially higher $M_{\mathrm{star}}$ than the pure LAEs. \textit{Right panels}:  Split contribution of SSP and DSP models to the age distribution of each family. It can also be appreciated that the DSP galaxies present the larger stellar masses.}
    \label{fig:mass_dist}
\end{figure*}

Regarding the stellar mass, the distribution presented in Fig.~\ref{fig:mass_dist} shows again a clear difference between pure LAEs (\mbox{median $M_{\mathrm{star}}\sim5.6^{+12.0}_{-5.5}\times10^{8}\ M_{\odot}$}) and \mbox{no-Ly$\alpha$} LBGs (median \mbox{$M_{\mathrm{star}}\sim3.5\pm1.1\times10^{9}\ M_{\odot}$}), with $M_{\mathrm{star}}$ an order of magnitude higher for the latter ones. Notice also that the \mbox{LAE-LBGs} present a smooth $M_{\mathrm{star}}$ distribution along a wider $M_{\mathrm{star}}$ range, with a median value of \mbox{$2.3\pm1.7\times10^{9}\ M_{\odot}$}. Here again, when we separate the contribution of SSP and DSP models, it can be appreciated that the more massive side of the $M_{\mathrm{star}}$ distribution is driven by the DSP sources while the low side of the mass distribution corresponds to the SSP ones. This result is not surprising once we take into account that, according with Fig.~\ref{fig:age_dist}, the DSP galaxies are much older and therefore have typically been forming stars for a much longer time, becoming more massive on average. Additionally, the detection of a relevant second SP is only possible in the most massive galaxies, where the old population presents a strong enough brightness at the longest wavelengths. In the particular case of the \mbox{LAE-LBGs}, these DSP objects could be understood as \mbox{no-Ly$\alpha$} LBGs that see their star formation increased by some triggering physical mechanism (mergers or neighbour gravitational interaction, instabilities, large cosmic web gas accretion, etc.) becoming \mbox{LAE-LBGs} (see Fig.~\ref{fig:DSP_LAEs_LBGs}).

\subsection{Burst strength}
\begin{figure}
	\includegraphics[width=\columnwidth]{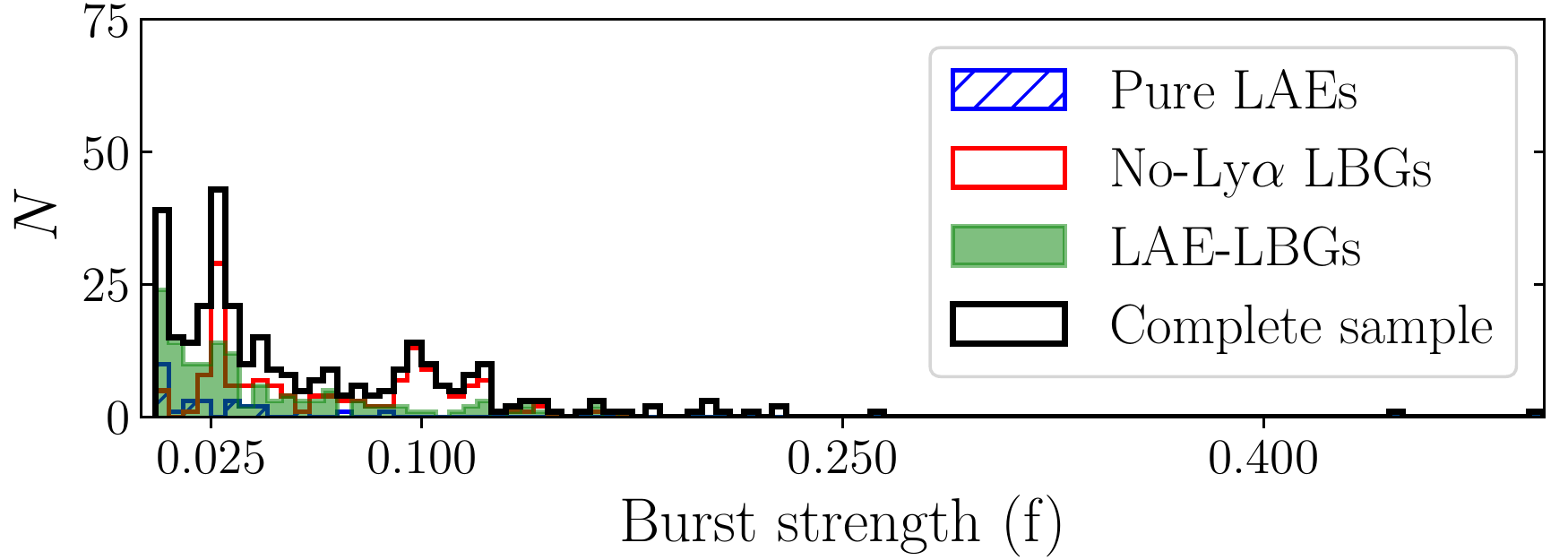}
   \caption{Distribution of the burst strength in the sources better approached by DSP models. The higher frequencies at low burst strength value indicate that even though two SPs are needed to properly model these objects, their SEDs are mostly the product of a main massive dominant old SP, with the young SP being almost irrelevant if not for the UV luminosity.}
    \label{fig:f_bursts}
\end{figure}

For the 326 galaxies better fitted using DSP models, a study of the relevance of each population in terms of mass is done by looking at the burst strength parameter $f$, whose distribution is shown in Fig.~\ref{fig:f_bursts}. We find that the burst strength remains low in almost all cases, with a 96\% of the objects at \mbox{$f<0.17$}. This distribution shows that even in the cases where a DSP modelling gives better results, these galaxies are still dominated by the main old SP in terms of stellar mass. Furthermore, the $M_{\mathrm{star}}$ of the young SP is almost negligible. However, it is important to highlight that the relevance of this young population comes with the conspicuous Ly$\alpha$ emission line and UV luminosity, which could not be reproduced using only a single old SP. No significant differences of the burst strength distribution were noticed among the three observational subgroups and no trend with redshift was neither found.

\subsection{Stellar mass functions}
\label{sec:SMF_res}
With the stellar masses derived from the \mbox{best-fitting} SP models we build Stellar Mass Functions (SMFs) at each redshift up to our stellar mass completeness. To estimate this completeness $M_{\mathrm{star},\mathrm{lim}}$, we make use of the technique employed in \textit{e.g.}, \citet{Pozzetti2010} and \citet{Davidzon2017} to calculate the stellar mass limit for a survey limited in magnitude. This method consists on taking the masses derived in each redshift bin and rescaling them to the magnitude limit of our survey:

\begin{equation}
\label{eq:M_resc}
 \log(M_{*,\mathrm{resc}})=\log(M_{\mathrm{star}})+0.4(m-m_{\mathrm{lim}}).
\end{equation}\

\noindent We adopt \mbox{$m_{\mathrm{lim}}\sim27$ AB} as an approximation of the average 3$\sigma$ limit detection in the SHARDS filters. The $M_{\mathrm{star},\mathrm{lim}}$ is then defined as the 90th percentile of the $M_{\mathrm{star},\mathrm{resc}}$ distribution. With this method, we estimate an average \mbox{$M_{\mathrm{star},\mathrm{lim}}\sim7.4\times10^{9}\ M_{\odot}$} for our stellar mass sample. This $M_{\mathrm{star}}$ completeness means that our SMFs are dominated by the LBGs population, as the majority of pure LAEs present masses below that limit (see Fig.~\ref{fig:mass_dist}). Additionally, a $V_{\mathrm{max}}$ correction \citep{Schmidt1968} is considered when building our SMFs. The main advantage of the $V_{\mathrm{max}}$ correction is that it directly provides the normalisation of the SMF. To model the SMF, we make use of the widely used \citet{Schechter1976} function:

\begin{equation}
\label{eq:Sch_mass}
 \phi(M)dM=\phi^{*}\exp(-M/M^{*})(M/M^{*})^{\alpha}dM/M^{*},
\end{equation}\

\noindent which can be better expressed in the $\log M$ space when working with SMFs:

\begin{align}
\label{eq:Sch_logmass}
 \phi(M)d\log M=\phi^{*}\ln10\exp(-10^{\log M-\log M^{*}})\notag\\
 \times\ (10^{\log M-\log M^{*}})^{\alpha+1}d\log M.
\end{align}\

\begin{figure*}
	\includegraphics[width=\textwidth]{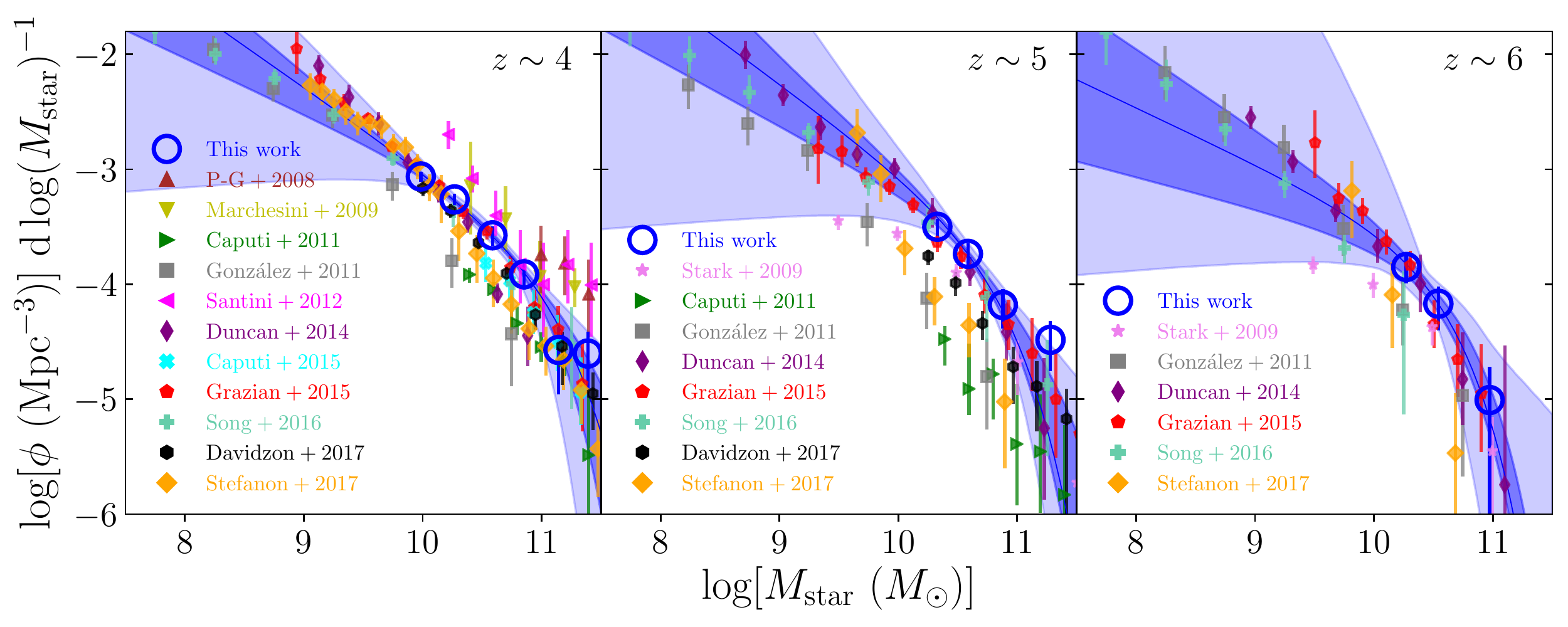}
   \caption{SMF at \mbox{$z\sim4$}, \mbox{$z\sim5$} and \mbox{$z\sim6$}. The Monte Carlo best fit is indicated with the solid blue line, while the darker and lighter blue contours correspond to the 68\% and 99.7\% confidence intervals, respectively. For comparison, we also show previous SMF calculations from \citet{PerezGonzalez2008} (\mbox{$3.5<z<4.0$}), \citet{Marchesini2009} (\mbox{$3.0<z<4.0$}), \citet{Stark2009}, \citet{Caputi2011}, \citet{Gonzalez2011}, \citet{Santini2012}, \citet{Duncan2014}, \citet{Caputi2015}, \citet{Grazian2015}, \citet{Song2016}, \citet{Davidzon2017} and \citet{Stefanon2017}. All SMFs have been rescaled to a Salpeter IMF for comparison.}
    \label{fig:SMFs}
\end{figure*}

The resulting SMFs are shown in Fig.~\ref{fig:SMFs} as well as the best Schechter fits and their $1\sigma$ and $3\sigma$ confidence intervals derived from Monte Carlo simulations perturbing the points themselves as well as the $M_{\mathrm{star}}$ bin sizes and centres, as explained in more detail in appendix~\ref{sec:appendix}. The error bars of our points correspond to the Poissonian uncertainties. The few points available at \mbox{$z\sim5$-6} up to our stellar mass completeness make the estimation of the $\alpha$ slope very difficult at these $z$, hence the large uncertainties derived from the fitting. Moreover, the calculated \mbox{$M_{\mathrm{star},\mathrm{lim}}\sim8.3\times10^{9}\ M_{\odot}$} at \mbox{$z\sim6$} seems to be slightly underestimated as it includes incomplete SMF points close to the $M^{*}$ knee for certain perturbations of the $M_{\mathrm{star}}$ bins, which derive in positive $\alpha$ values. To avoid this issue, we constrain the $\alpha$ fitting to \mbox{$-3.0<\alpha<-0.9$} at \mbox{$z\sim6$} and limit the SMF points to those strictly increasing up to our $M_{\mathrm{star},\mathrm{lim}}$ as we consider that further ones are actually incomplete. Note that this approach could be slightly biasing the obtained $\alpha$ values towards more negative values at \mbox{$z\sim6$}. More details on the fitting process, as well as the significance contours of the fitted Schechter parameters at each redshift are given in appendix~\ref{sec:appendix}.

\begin{figure}
	\includegraphics[width=\columnwidth]{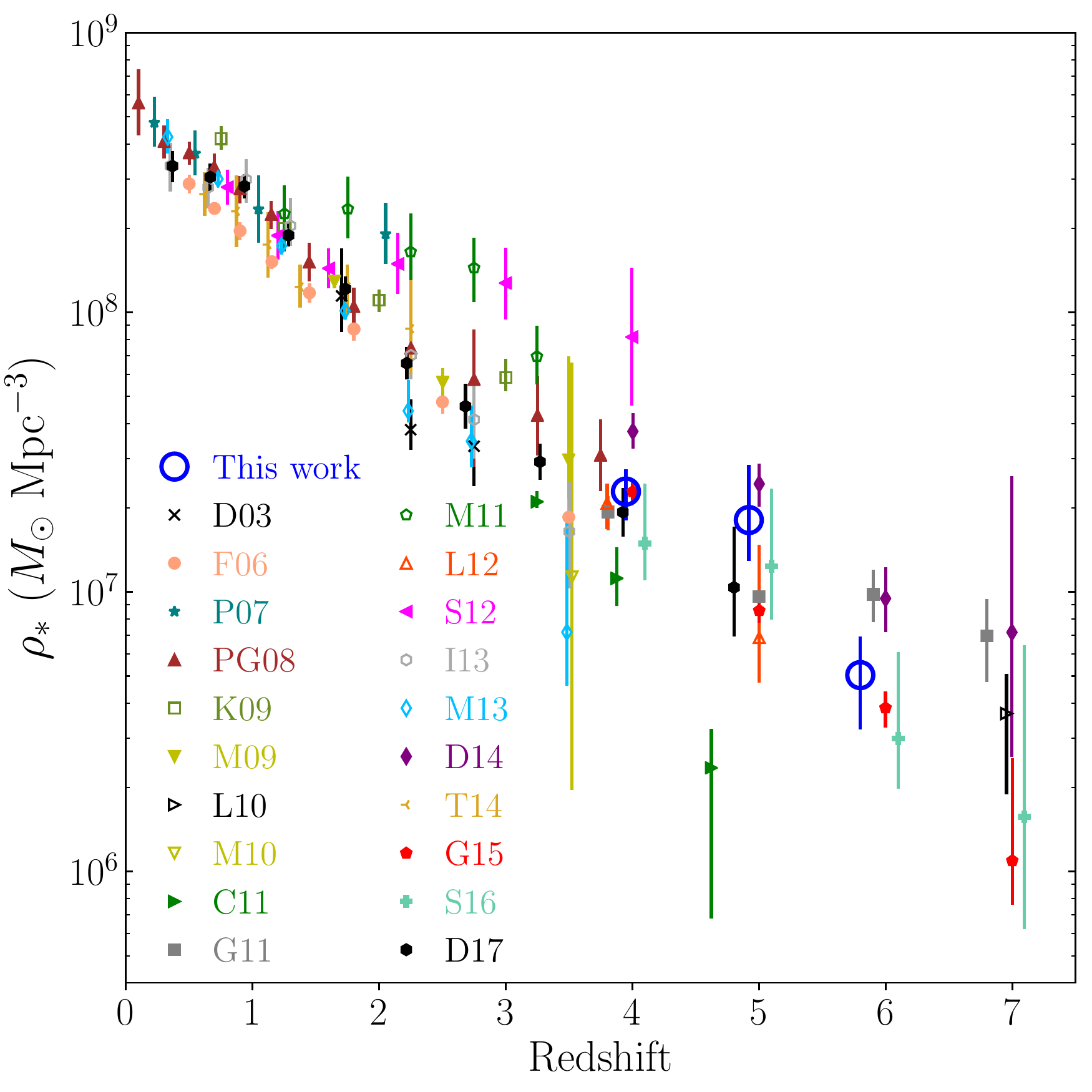}
   \caption{SMD obtained by integration of the SMFs at each $z$. To give a global view of the SMD evolution with $z$, we show previous estimations from \citet[][D03]{Dickinson2003}, \citet[][F06]{Fontana2006}, \citet[][P07]{Pozzetti2007}, \citet[][PG08]{PerezGonzalez2008}, \citet[][K09]{Kajisawa2009}, \citet[][M09]{Marchesini2009}, \citet[][M10]{Marchesini2010}, \citet[][L10]{Labbe2010}, \citet[][C11]{Caputi2011}, \citet[][G11]{Gonzalez2011}, \citet[][M11]{Mortlock2011}, \citet[][L12]{Lee2012}, \citet[][S12]{Santini2012}, \citet[][I13]{Ilbert2013}, \citet[][M13]{Muzzin2013}, \citet[][D14]{Duncan2014}, \citet[][T14]{Tomczak2014}, \citet[][G15]{Grazian2015},
   \citet[][S16]{Song2016} and 
   \citet[][D17]{Davidzon2017}. All SMDs are rescaled to a Salpeter IMF.}
    \label{fig:SMD}
\end{figure}

By integrating the SMF from $10^{8}$ to \mbox{$10^{13}$ $M_{\odot}$} \citep[as in, \textit{e.g.},][]{Duncan2014, Grazian2015, Song2016}, we obtain the Stellar Mass Density (SMD) at each redshift. The \mbox{best-fitting} Schechter parameters are summarised in Table~\ref{tab:Sch_par} as well as the calculated SMDs. A comparison of our SMD estimations with previous works is shown in Fig.~\ref{fig:SMD}, where it can be appreciated that our calculations follow the general trend with $z$ reported by previous authors, presenting a especially large uncertainty at \mbox{$z\sim5$-6} due to the mentioned lack of information in the low mass regime of our SMF at that redshift and the corresponding uncertainty of the estimated $\alpha$ slope.

\begin{table}
    \centering
     \caption{SMF \mbox{best-fitting} Schechter parameters and the corresponding SMD obtained by integration of the SMF from $10^{8}$ to \mbox{$10^{13}$ $M_{\odot}$}. The shown uncertainties correspond to 1$\sigma$ significance. The redshift ranges represented correspond to \mbox{$3.5\leq z<4.5$}; \mbox{$4.5\leq z<5.5$} and \mbox{$5.5\leq z<6.5$}, respectively.}
    \label{tab:Sch_par}
    \begin{tabular}{ccccc}
    \hline
        $\langle z\rangle$   &   $\log M^{*}$  &   $\log\phi^{*}$   &   $\alpha$ & $\log\rho_{*}$\\
\hline
  $4$ & $11.06^{+0.33}_{-0.27}$ & $-4.14^{+0.39}_{-0.45}$ & $-1.72^{+0.24}_{-0.14}$ & $7.36^{+0.08}_{-0.10}$\\
  $5$ & $10.78^{+0.53}_{-0.07}$ & $-3.97^{+0.10}_{-0.79}$ & $-1.76^{+0.19}_{-0.26}$ & $7.26^{+0.20}_{-0.15}$\\
  $6$ & $10.51^{+0.08}_{-0.03}$ & $-4.06^{+0.02}_{-0.06}$ & $-1.49^{+0.22}_{-0.21}$ & $6.70^{+0.14}_{-0.19}$\\
 \hline
    \end{tabular}
\end{table}

\subsection{SFR-\texorpdfstring{$\boldsymbol{M_{\mathrm{star}}}$}{} relation}
\label{sec:SFR-M}
\begin{figure*}
	\includegraphics[width=\textwidth]{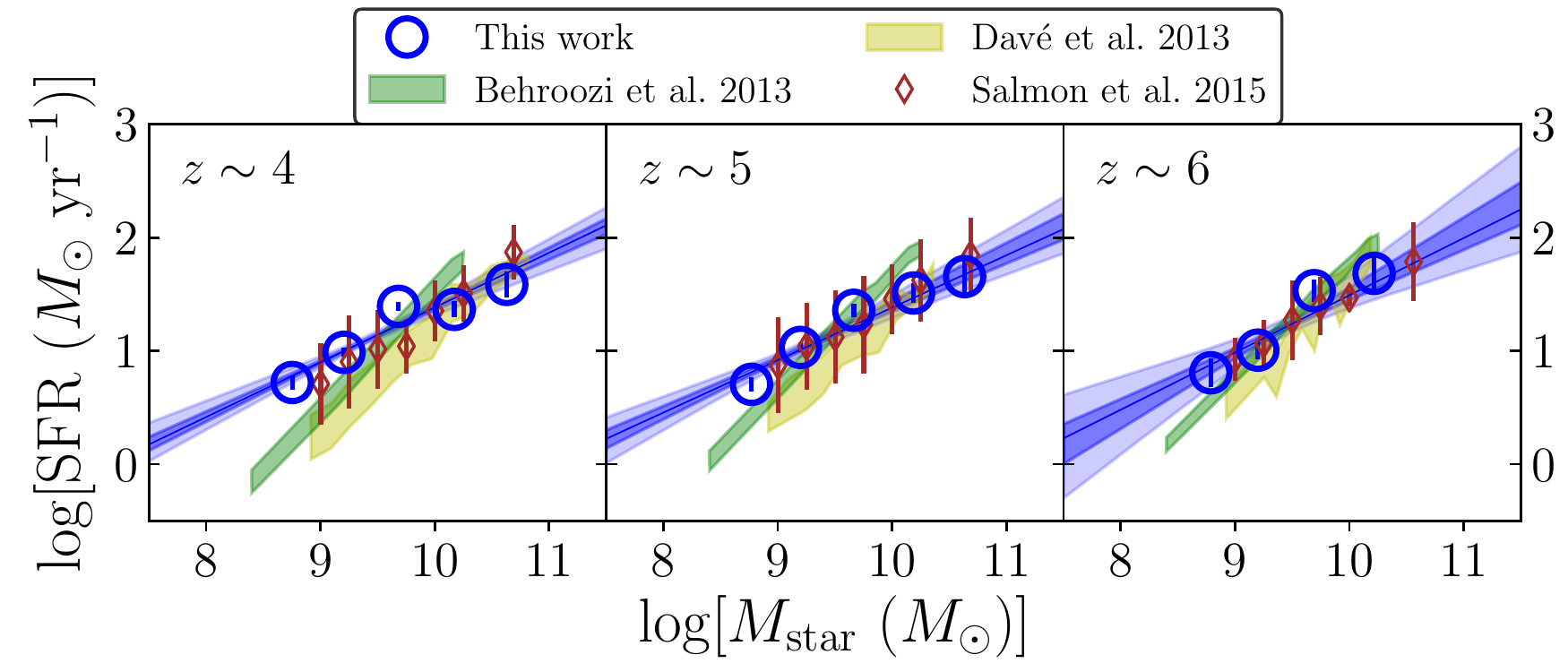}
   \caption{\mbox{SFR-$M_{\mathrm{star}}$} relation measured at each redshift using the \mbox{UV-derived} SFRs. Pure LAEs are not considered in this \mbox{SFR-$M_{\mathrm{star}}$} relation, since their SFRs are only estimated through their Ly$\alpha$ luminosity. The error bars of our data (blue circles) is associated to the standard error of the median SFR at each $M_{\mathrm{star}}$ bin. The blue solid line corresponds to the best fit and the darker and lighter blue regions delimit the 1$\sigma$ and 3$\sigma$ confidence intervals of the fit, respectively. Theoretical predictions from \citet[faded green contour]{Behroozi2013} and \citet[faded yellow contour]{Dave2013} show slightly steeper slopes than the observations from \citet{Salmon2015} and our sample. The slope of the \mbox{SFR-$M_{\mathrm{star}}$} relation does not show a significant change with $z$.}
    \label{fig:SFR-M}
\end{figure*}

To build \mbox{SFR-$M_{\mathrm{star}}$} relations at each $z$ and compare them with previous estimations, we make use of the SFRs calculated for this sample in \citet{ArrabalHaro2018} using the \citet{Kennicutt1998} and \citet{Madau1998} prescriptions after correcting for both galactic and internal dust extinction following \citet{Schlafly2011} and \citet{Calzetti2000}, respectively. Those SFRs were calculated using the Ly$\alpha$ emission line for the pure LAEs and the UV luminosity for the LBGs. The SFRs of the pure LAEs are estimated through their Ly$\alpha$ luminosity in \citet{ArrabalHaro2018}, thus they are not on equal terms with the \mbox{UV-derived} \mbox{SFR-$M_{\mathrm{star}}$} relation. To avoid using multiple different SFR indicators, we only make use of the SFRs estimated through $L_{1500}$, which allow us to compare our results with previous works studying this relation in a similar fashion such as, \textit{e.g}, \citet{Salmon2015}. Moreover, the SFRs derived through the Ly$\alpha$ line luminosity can be strongly affected by resonant scattering. For a recent detailed study of the \mbox{SFR-$M_{\mathrm{star}}$} relation estimating SFRs of \mbox{high-$z$} LAEs through their Ly$\alpha$ luminosity, see, \textit{e.g.}, \citet{Santos2020}. The \mbox{so-called} \mbox{SFR-$M_{\mathrm{star}}$} main sequence (Fig.~\ref{fig:SFR-M}) is modelled using the common linear approach between the logarithm of these magnitudes \citep[\textit{e.g.},][]{Salmon2015}:

\begin{equation}
\label{eq:SFR-M}
 \log[\mathrm{SFR}\ (M_{\odot}\ \mathrm{yr^{-1}})]=\beta\log[M_{\mathrm{star}}\ (M_{\odot})] + C.
\end{equation}\

The error bars shown in Fig.~\ref{fig:SFR-M} correspond to the standard error of the median SFR at each $M_{\mathrm{star}}$ bin. To fit the slope and zero point of the relation we applied Monte Carlo methods not only perturbing the points within the errors, but also the $M_{\mathrm{star}}$ bin centres and sizes in a range of 0.1-0.5 dex. To compare our results, we use data from semi-empirical models \citep{Behroozi2013}, hydrodynamic simulations \citep{Dave2013} and observational data from \citet{Salmon2015}. Note that the distribution of these observational data matches very well ours, though the trend of the theoretical models is slightly steeper. The best-fitted values of $\beta$ and $C$ are given in Table~\ref{tab:SFR-M_fits}. It can be noticed that the slope of the \mbox{SFR-$M_{\mathrm{star}}$} remains invariable with $z$ within errors, as it has been previously reported in the literature \citep{Stark2009, Gonzalez2010, Papovich2011, Salmon2015}. The implications of this absence of evolution in the \mbox{SFR-$M_{\mathrm{star}}$} relation with $z$ will be further discussed in Sec~\ref{sec:Discussion}.

\begin{figure*}
	\includegraphics[width=\textwidth]{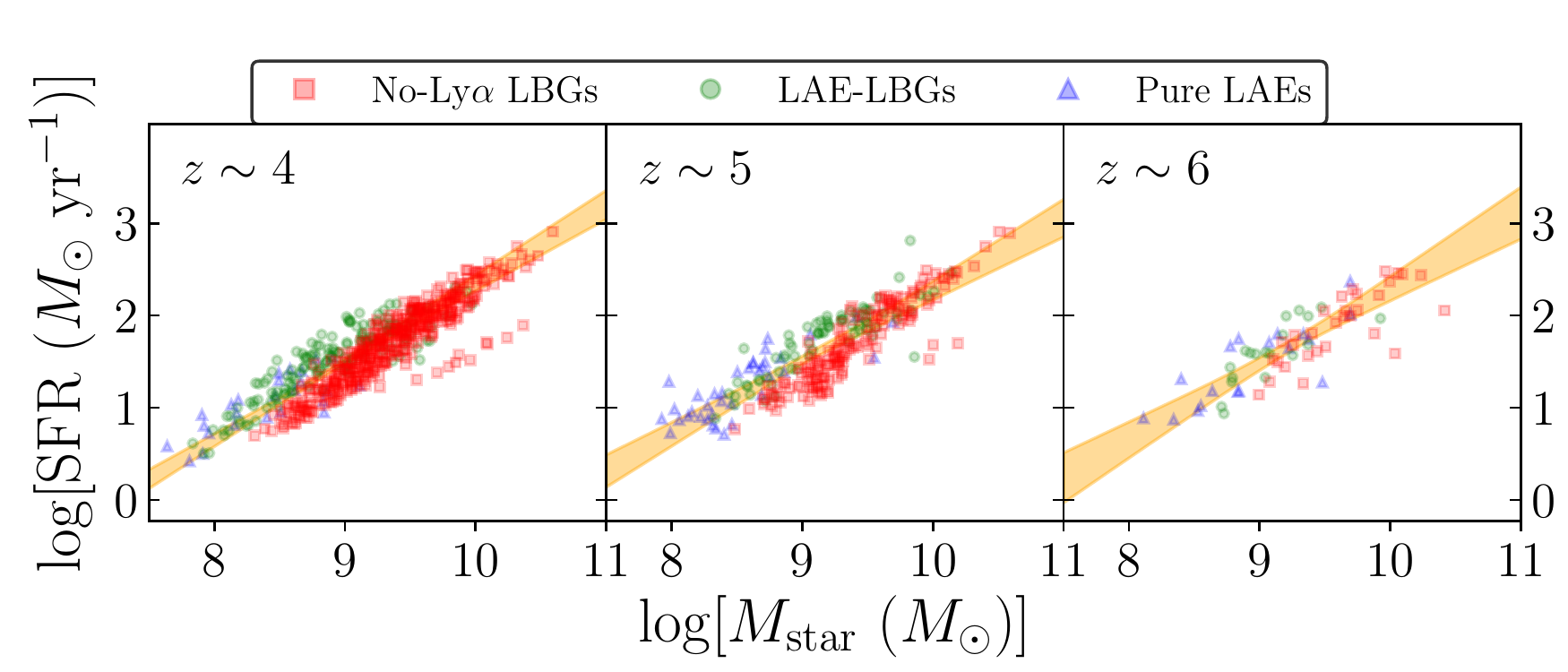}
   \caption{\mbox{SFR-$M_{\mathrm{star}}$} main sequence measured at each redshift using the SFRs given by the \mbox{best-fitting} SSP {\sc cigale} models. The orange region delimit the 3$\sigma$ confidence interval of the fit. The slope of the \mbox{SFR-$M_{\mathrm{star}}$} relation does not change with $z$ within our redshift range.}
    \label{fig:SFR-M_cigale}
\end{figure*}

In order to study the distribution of the different subclasses along this main sequence employing the same SFR estimator for all of them, we make use of the main sequence built up with the SFRs derived from the \mbox{best-fitting} SSP {\sc cigale} models (Fig.~\ref{fig:SFR-M_cigale}). Note that the different nature of the SFR estimators makes their absolute values difficult to compare as they can differ substantially. Indeed, as reported in, \textit{e.g.}, \citet{OtiFloranes2010}, usual SFR estimators tend to have strong assumptions on the SFH. These assumptions can be incorrect when using \mbox{burst-like} SFHs as the ones employed for \mbox{high-$z$} galaxies in this paper, resulting in very differing SFR estimations. Because of this, the absolute \mbox{model-derived} SFRs from \mbox{burst-like} SPs are higher than those from \citet{Behroozi2013} and \citet{Dave2013}, who did not employ short-lived star formation episodes. Nevertheless, their main sequence slope is similar to the \mbox{model-derived} slope obtained for our sample. For more details on differences in SFR calculation, see \citet{Boquien2014} or \citet{Boquien2016}, among others. The reason to study the \mbox{SFR-$M_{\mathrm{star}}$} using the SFRs from the models is, apart from having an additional measure of the main sequence slope, to analyse relative differences between subclasses from a common SFR estimation. The \mbox{best-fitting} parameters of this \mbox{model-derived} \mbox{SFR-$M_{\mathrm{star}}$} relation are also shown in Table~\ref{tab:SFR-M_fits}. The slope obtained in this second case is slightly steeper than those obtained employing the \mbox{UV-derived} SFRs, getting closer to the theoretically predicted by \citet{Behroozi2013} and \citet{Dave2013}. Furthermore, the \mbox{SFR-$M_{\mathrm{star}}$} relation derived from the \mbox{best-fitting} SSP models also remains constant between \mbox{$z=4$-6}, reinforcing this result.

Regarding the different subgroups distribution, it can be appreciated that pure LAEs occupy the left end of the main sequence, corresponding to lower stellar masses and SFRs, while LBGs conform the bulk at intermediate and large masses. Note that both \mbox{LAE-LBGs} and pure LAEs appear in the upper side of the main sequence, in agreement with the idea of these sources experimenting a recent star-forming episode, while \mbox{no-Ly$\alpha$} LBGs are placed in the bottom-middle side of it. 

\begin{table}
    \centering
     \caption{\mbox{Best-fitting} parameters for the \mbox{SFR-$M_{\mathrm{star}}$} main sequence built up using \mbox{UV-derived} SFRs (second and third columns) and \mbox{model-derived} SFRs (last two columns).}
    \label{tab:SFR-M_fits}
    \begin{tabular}{ccccc}
    \hline
        $\langle z\rangle$ & $\beta_{L_{1500}}$ & $C_{L_{1500}}$ & $\beta_{\mathrm{mod}}$ & $C_{\mathrm{mod}}$\\
\hline
  $4$ & $0.48^{+0.07}_{-0.10}$ & $-3.45^{+0.92}_{-0.70}$ & $0.83^{+0.09}_{-0.06}$ & $-6.01^{+0.52}_{-0.76}$\\
  $5$ & $0.46^{+0.12}_{-0.10}$ & $-3.25^{+0.91}_{-1.13}$ & $0.79^{+0.10}_{-0.11}$ & $-5.56^{+0.92}_{-0.97}$\\
  $6$ & $0.51^{+0.26}_{-0.19}$ & $-3.56^{+1.80}_{-2.52}$ & $0.82^{+0.15}_{-0.16}$ & $-5.89^{+1.42}_{-1.43}$\\
 \hline
    \end{tabular}
\end{table}

\section{Discussion}
\label{sec:Discussion}

One of the particularities of this study lays on the very short $\tau$ values found for the \mbox{best-fitting} models of the sample. These short $\tau$ values represents bursts of star formation. This result leads to younger ages than those typically obtained in previous models of  \mbox{high-$z$} LAEs and LBGs using fairly constant SFHs, where the estimate ages are of the order of few hundreds Myr \citep[see, \textit{e.g},][]{Dayal2012}. However, shorter ages like the ones presented here for SSP star-forming sources have also been obtained for LAEs and LBGs using different SFHs. \citet{Jiang2016}, for example, found a similar age bimodality (also present in their $M_{\mathrm{star}}$ distribution) when modelling \mbox{high-$z$} galaxies with exponentially declining SFHs with larger \mbox{$e$-folding} time (\mbox{$\tau$=200 Myr}) and smoothly raising SFHs. Employing a constant SFH, \citet{Yuma2010} also found very short ages (median age of \mbox{25 Myr}) for \mbox{$z\sim5$} LAEs. In any case, the age estimation of \mbox{high-$z$} LAEs and LBGs through SED fitting presents large uncertainties independently of the input parameters of the models employed, and so our interest is in spotting relative differences between observational subfamilies rather than calculate exact absolute age values.

The stellar masses found for \mbox{high-$z$} LAEs and LBGs are better constrained in the literature, presenting values in the \mbox{10$^{8}$-10$^{11}$ $M_{\odot}$} range with median values of a few \mbox{$10^{9}$ $M_{\odot}$} \citep[see, \textit{e.g},][]{Yuma2010, Dayal2012, Duncan2014, Grazian2015, Jiang2016, Song2016, Davidzon2017}, with LAEs typically presenting lower masses, which can also be related with a selection bias effect as stated in \citet{Dayal2012}. The stellar masses found in this work are in good agreement with previous estimations at high redshifts.

\subsection{LAEs and LBGs stellar population differences}
\label{sec:SP_disc}
From the ages and stellar masses derived with {\sc cigale} we can build some relations between the different observational families previously defined. As highlighted in Sec.~\ref{sec:Results}, the age values are on the low side due to the existing degeneracy between dust extinction, metallicity and the age itself, and so these absolute values should be taken with care. Nevertheless, we can use them to trace age differences between our various families of LAEs and LBGs.

\begin{figure}
	\includegraphics[width=\columnwidth]{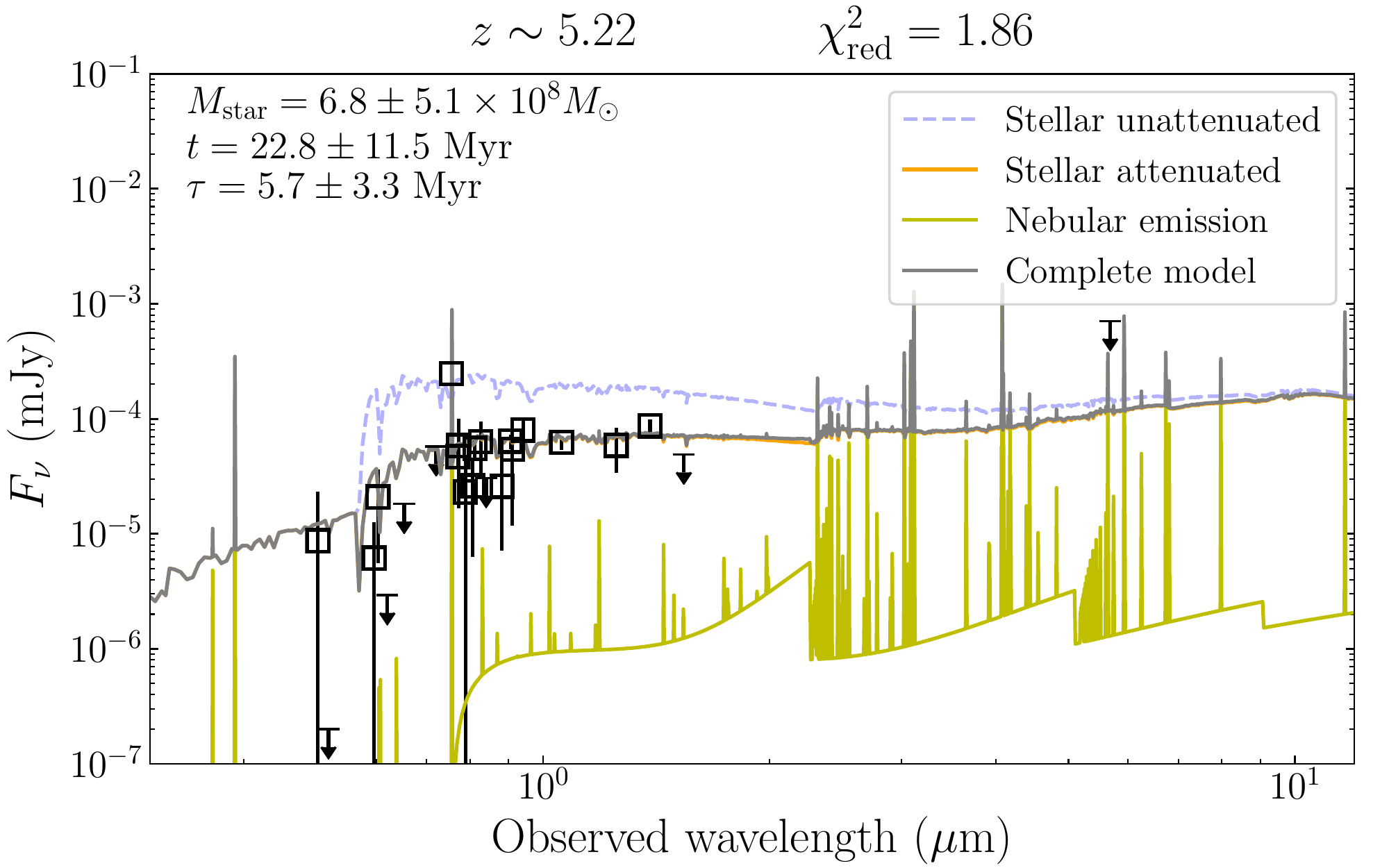}
   \caption{Best model for the pure LAE SHARDS J123640.20+621228.8. The black squares are the photometric points used in the fit. The final model (grey line) is split into the attenuated stellar emission (orange line) and the nebular emission (yellow line). The \mbox{non-attenuated} stellar emission is also represented by the dashed bluish line. The SED presents a clear Ly$\alpha$ emission line, but the absence of strong continuum points at the longest sampled wavelengths makes it possible to fit it with a single young and not so massive SP.}
    \label{fig:SSP_pure_LAE}
\end{figure}

On the one side, we have the pure LAEs, defined as objects with strong Ly$\alpha$ line emission but a faint UV continuum (\mbox{$m_{1500}\gtrsim27$ AB}). These sources typically present low stellar masses in their young SP (median \mbox{$M_{\mathrm{star}}=5.6^{+12.0}_{-5.5}\times10^{8}\ M_{\odot}$}). Additionally, the presence of strong Ly$\alpha$ emission quickly decays with time as it traces the Lyman continuum radiation which is only produced by O and late type B stars with \mbox{$M>10\ M_{\odot}$} and lifetimes of a few Myr, and so it is an indicator of recent star formation. That, on top of the low $M_{\mathrm{star}}$, indicates that these  galaxies should be typically young, since they are currently in a star-forming burst but have not been forming stars long enough, in the past, to present larger stellar masses. This hypothesis is confirmed by the median age obtained with the {\sc cigale} fitting for this family (\mbox{$26^{+41}_{-25}$ Myr}). According to this, many of these sources could indeed be experimenting one of their first episodes of star formation, or at least one strong enough to overtake all the older stars in luminosity. Furthermore, we also find that most of these objects can be explained using a single decaying exponential SFH. An example of this is shown in Fig.~\ref{fig:SSP_pure_LAE}, where the Ly$\alpha$ emission line, together with the absence of strong UV continuum photometric points makes it possible to model these pure LAEs in terms of a single young and low mass SP. Note that some of these SSP-fitted sources could actually host an old SP from previous star-forming episodes. However, these old SPs are not massive enough to be identified over the young SP luminosity. The 27 DSP pure LAEs found in this work can be modelled as older and more massive galaxies that are experimenting a recent star-forming episode. This recent star formation is not strong enough to raise the UV continuum up to a flux level measurable in SHARDS, but it does raise the Ly$\alpha$ emission associated to young short-living stars, while the old SP makes them detectable at longer wavelengths in the IRAC range.

\begin{figure*}
	\includegraphics[width=\textwidth]{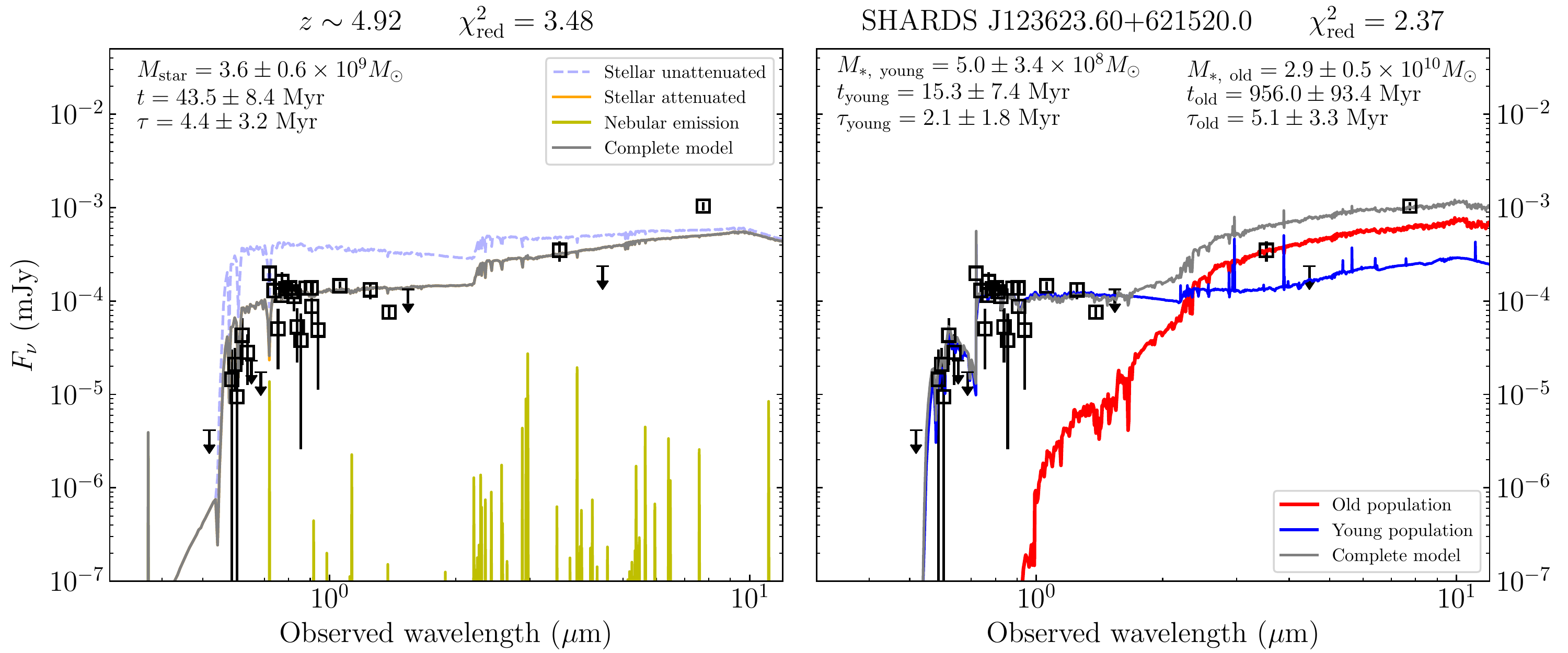}
    \vskip\baselineskip
    \includegraphics[width=\textwidth]{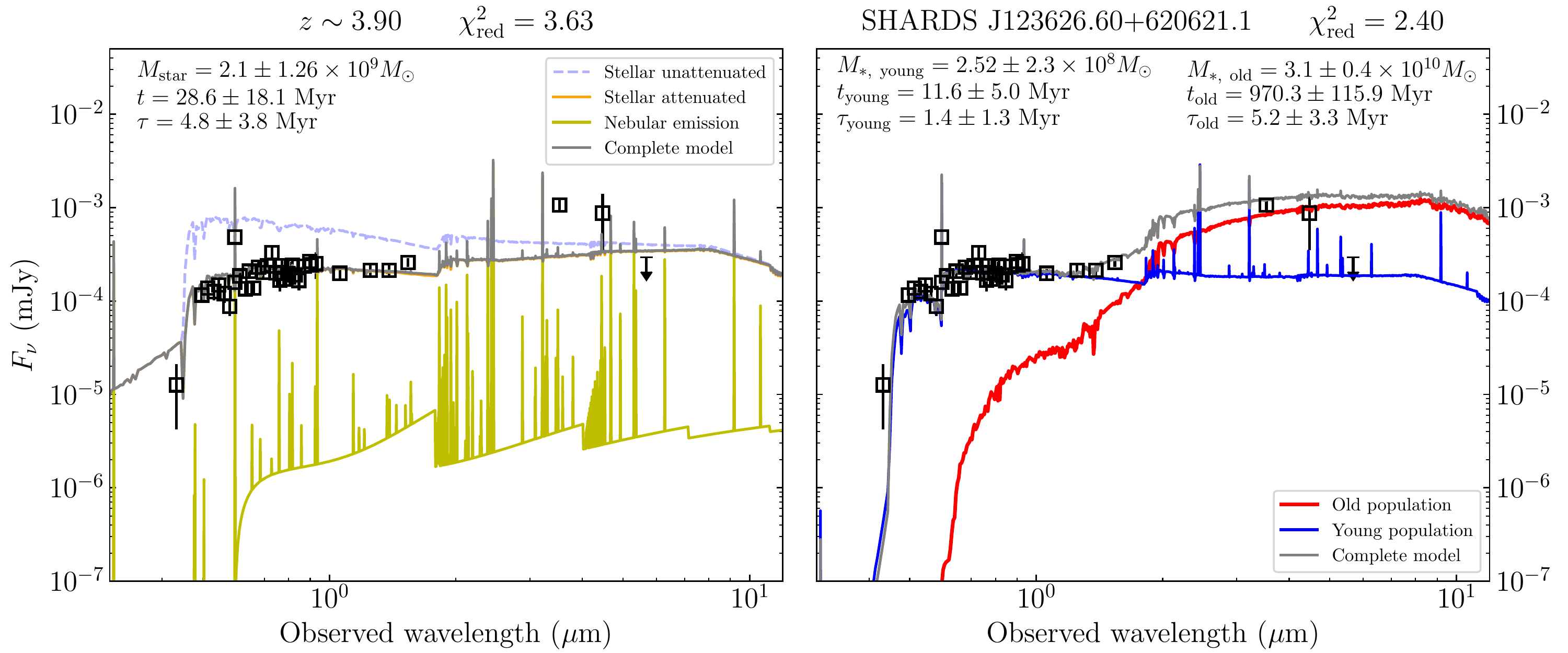}
 \caption{Best SSP and DSP solutions for two \mbox{LAE-LBGs} modelled each with two SPs. The objects SHARDS J123623.60+621520.0 and SHARDS J123626.60+620621.1 are shown in the upper and lower panels, respectively. In each case, the left frame shows the best SSP fit. The black squares are the photometric points of the SED, the yellow line shows the nebular emission, the stellar attenuated emission is represented in orange and the stellar \mbox{non-attenuated} emission is plotted with a dashed bluish line. The grey line shows the complete model. On the other side, the right frames correspond to the best DSP fits, where the different components of the emission have been omitted for clarity and the contributions of the old and young SPs are represented in red and blue, respectively. For the upper source, the best SSP model matches relatively well the continuum but is not able to reproduce the Ly$\alpha$ emission line detected. The addition of a second SP becomes necessary to not only to match that Ly$\alpha$ emission, but even to improve the continuum fit in the IRAC range. The opposite scenario can be seen in the lower source, where the best SSP fit manages to reproduce the line emission at the cost of leaving the reddest points unfitted. This time, the second population added is an old one that contributes to raise the continuum emission in the IRAC range.}
 \label{fig:DSP_LAEs_LBGs}
\end{figure*}

For the \mbox{LAE-LBGs}, the dichotomy found in their age and $M_{\mathrm{star}}$ distributions shown in the right panels of Figs.~\ref{fig:age_dist} and \ref{fig:mass_dist} set a similar differentiation between SSP \mbox{LAE-LBGs} and DSP \mbox{LAE-LBGs} as with pure LAEs. The objects of this family that can be fitted using a SSP are the youngest and less massive within the \mbox{LAE-LBGs}. These cases do not present any strong emission at longer wavelengths and can therefore be reproduced by a young SP (median age of \mbox{$27\pm6$ Myr}) with Ly$\alpha$ line in emission and a fairly flat UV continuum. According to the description given in the previous paragraph for the pure LAEs, the SSP \mbox{LAE-LBGs} would just be the most massive members of that same class: young galaxies well modelled by a SSP, with the only difference that SSP \mbox{LAE-LBGs} present a massive enough young SP (median \mbox{$M_{\mathrm{star}}=1.04\pm0.48\times10^{9}\ M_{\odot}$}) to show an UV continuum detectable in SHARDS, preventing the pure LAEs observational classification but actually belonging to the same kind of objects. On the other side, the DSP \mbox{LAE-LBGs} are much older and more massive. The need of a second SP to understand the SEDs of this subgroup comes from the presence of the Ly$\alpha$ emission line plus some bright IRAC points. Both emission features cannot be simultaneously fitted by a SSP with the characteristics employed in this work \citep[as in, \textit{e.g.},][]{RodriguezEspinosa2014}. In Fig.~\ref{fig:DSP_LAEs_LBGs}, we present two examples to illustrate what happens when we try to fit some of these objects SEDs with a SSP that either fits well the Ly$\alpha$ line but not the longer wavelength continuum or the other way around. The {\sc cigale} best solutions for this subgroup suggest that the DSP \mbox{LAE-LBGs} can be understood as older galaxies with an old SP that raises the continuum emission at longer wavelengths, currently experimenting a recent star-forming episode triggered by accretion of new gas or by mergers. It makes sense thinking about the observational definition of pure LAEs and \mbox{LAE-LBGs} being the same kind of galaxies (just LAEs) with the only difference that pure LAEs are more frequently fitted by SSP models, while \mbox{LAE-LBGs} require the addition of the old SP more often.

On the other side, we have the \mbox{no-Ly$\alpha$} LBGs, understood as galaxies selected through their Lyman break in the UV continuum emission, but without Ly$\alpha$ line detected in the SHARDS photometry (\mbox{$F(\mathrm{Ly}\alpha)\lesssim1.3^{+2.0}_{-1.3}\times 10^{-19}\ \mathrm{erg\ cm^{-2}}$}). This family is the oldest and most massive of the SSP models of the three predefined observational families, with a median stellar mass almost an entire order of magnitude above that of the pure LAEs (\mbox{$M_{\mathrm{star}}=3.5\pm1.1\times10^{9}\ M_{\odot}$}). The absence of a detectable Ly$\alpha$ line should not be taken as a secure indicator of the relative faintness of the most recent star formation in these objects, since the real Ly$\alpha$ emission of the galaxy can be strongly affected by dust extinction and resonant scattering through the interstellar medium. Indeed, dust extinction affects strongly both the Ly$\alpha$ line and the rest-frame UV continuum. Hence, there is an intrinsic selection effect towards galaxies with low internal extinction. Ly$\alpha$ photons scattering by neutral gas, on the other hand, plays an important role in the fraction of LBGs with and without Ly$\alpha$ emission. Regarding Ly$\alpha$ \ion{H}{i} resonant scattering, recent works estimate different escape fraction values depending on the galaxy population, as commented in Sec.~\ref{sec:Models}, from large \mbox{$f_{\mathrm{esc}}\sim0.5$} for bright \mbox{$z\sim2$-3} LAEs \citep{Sobral2018b} and typical LAEs \citep{Sobral2017, Sobral2019} to very low \mbox{$f_{\mathrm{esc}}\sim0.02-0.05$} for more massive and dusty \mbox{$z=2.23$} H$\alpha$ emitters \citep{Matthee2016}. In any case, these Ly$\alpha$ destruction or scattering phenomena are very difficult to quantify with our data so the discussion of this \mbox{no-Ly$\alpha$} LBGs family aims to give a general overview of the class, even though there could be particular cases not matching it.

Thus, the results from the {\sc cigale} fits point to the \mbox{no-Ly$\alpha$} LBGs typically being a more evolved stage after previous episodes of star formation. Thus these sources will have time to form a large amount of stars and therefore show larger stellar masses and stronger continuum emission supported by the old long-living stars. The need of a second SP to model this class is another problem difficult to answer. According to the idea of these more massive sources being the product of many previous star-forming episodes, it could also be thought that the most reasonable way of approaching them should be using more than one SP. However, the emission patterns that more easily differentiate young SP from old ones (as the nebular emission or the brightness of the UV region of the SED) are not conspicuous in these sources. This indicates that they do not have a really young SP (\mbox{$\lesssim25$ Myr}). Nonetheless, given that there exists a degeneracy in the number of relatively old SPs with different ages (understood as different star-forming episodes) these sources could  usually be reproduced by a single stellar population. As we adopted the BIC as a good indicator to estimate whether a second SP is needed in our models, the majority of situations where both the SSP and DSP approaches give similar $\chi^{2}$ solutions end up favouring the simplest model. Thus only a 16\% of the \mbox{no-Ly$\alpha$} LBGs do need the extra SP. Furthermore, since the UV continuum close to Ly$\alpha$ is driven by the most recent SF episodes the SEDs of these sources are most of the time (80\%) fitted by SSPs in the range of \mbox{30-150 Myr}, still older than those typically fitted to the pure LAEs and the SSP \mbox{LAE-LBGs}, but certainly younger than what they could be if they were actually hosting an underlying old SP. What we want to emphasise here is that even though those objects are well modelled by a SSP of the nature described above, though there could be some other faint and much older extra SPs. As we are adopting the simplest model in these situations, we should be aware of a possible bias towards younger ages for this particular family of \mbox{no-Ly$\alpha$} LBGs.

\begin{figure}
	\includegraphics[width=\columnwidth]{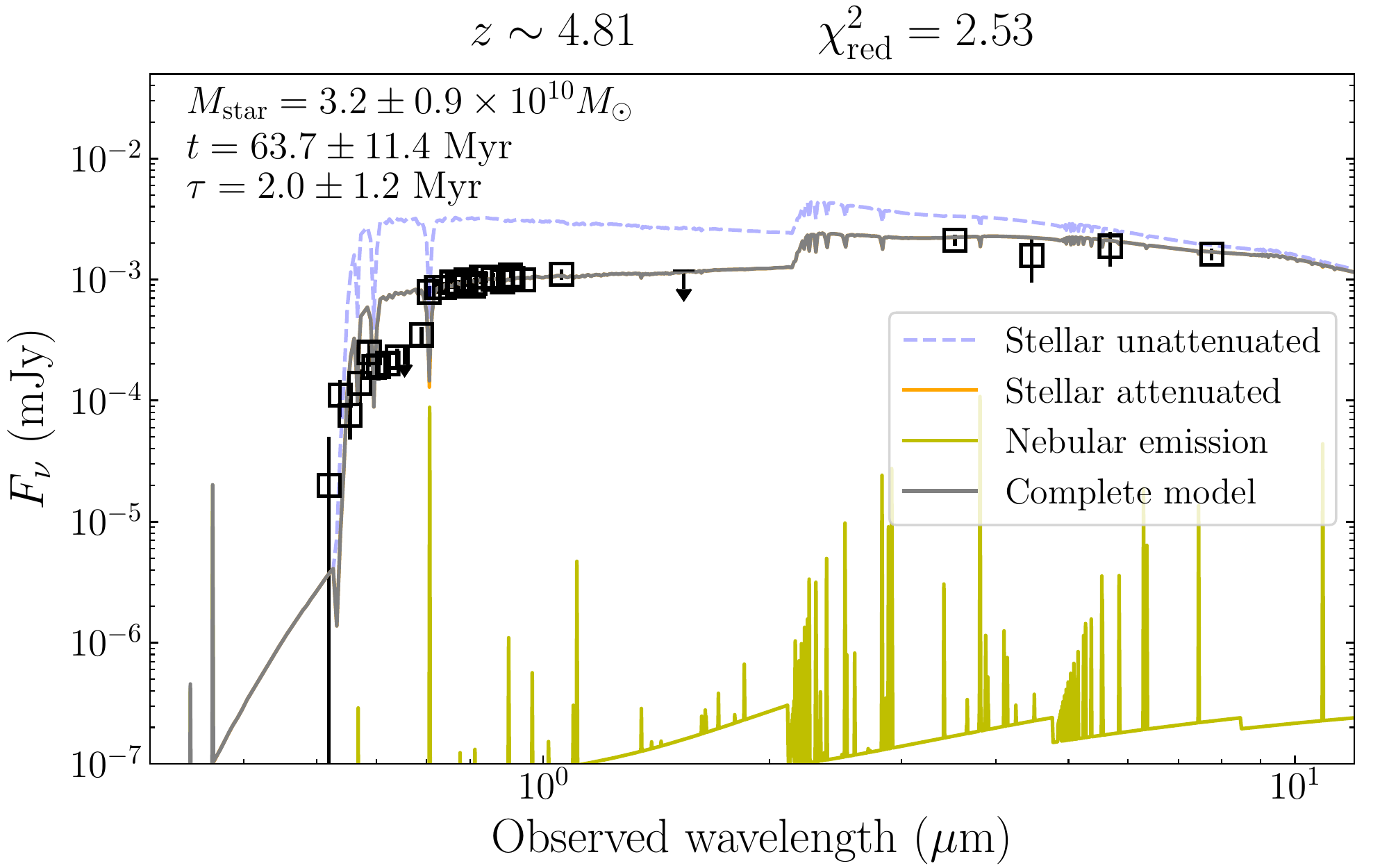}
   \caption{SSP model taken as the best solution for the \mbox{no-Ly$\alpha$} LBG SHARDS J123757.50+621718.7. The absence of Ly$\alpha$ emission line makes it possible to fit the continuum of the SED even up to the IRAC measurements using a single relatively old, massive stellar population. Note that the nebular emission is so low that the complete spectrum practically corresponds to the attenuated stellar emission.}
    \label{fig:SSP_LBG}
\end{figure}

\begin{figure}
	\includegraphics[width=\columnwidth]{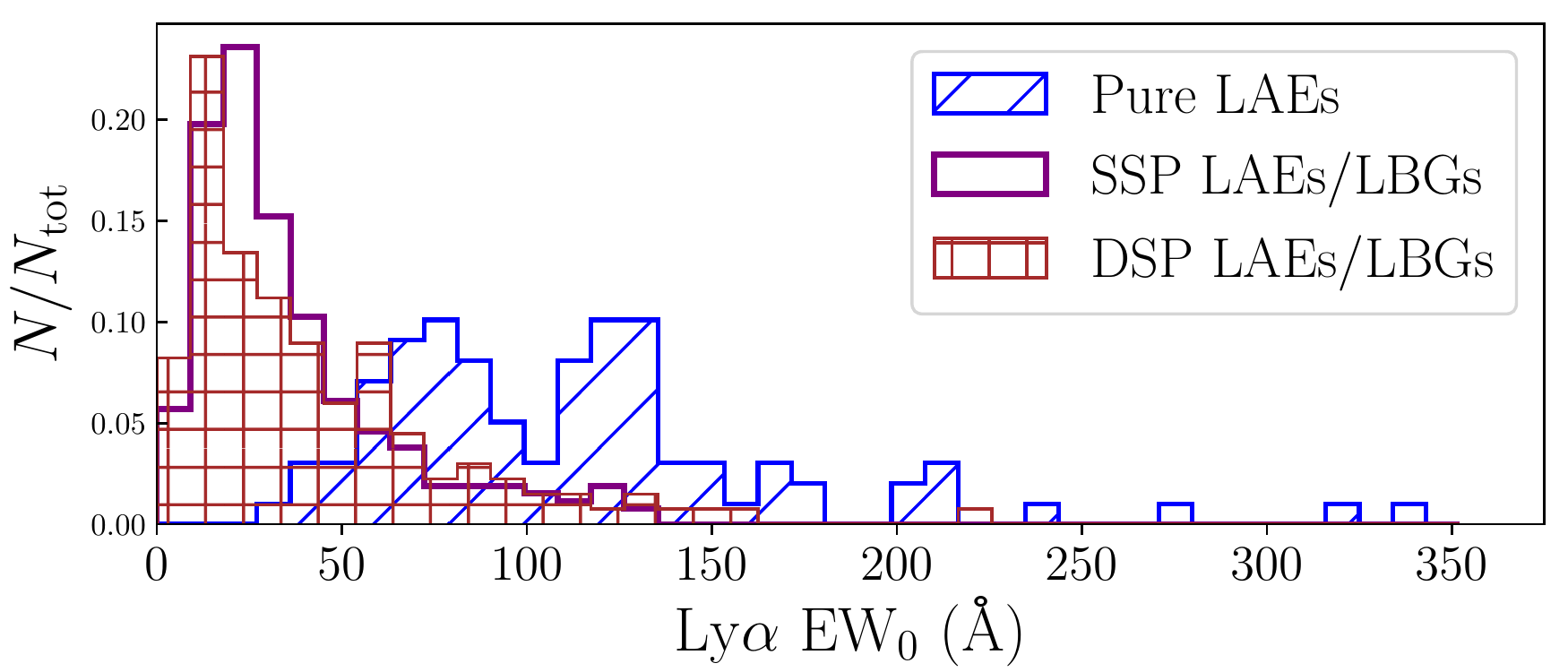}
   \caption{Rest-frame Ly$\alpha$ EW distribution of pure LAEs (blue diagonals), SSP \mbox{LAE-LBGs} (unfilled purple) and DSP \mbox{LAE-LBGs} (brown squares) weighted by the total amount of objects belonging to each subclass. No relation is found between Ly$\alpha$ EW and the need of a second SP when modelling the SEDs of \mbox{LAE-LBGs}.}
    \label{fig:EW_dist}
\end{figure}

\begin{figure}
	\includegraphics[width=\columnwidth]{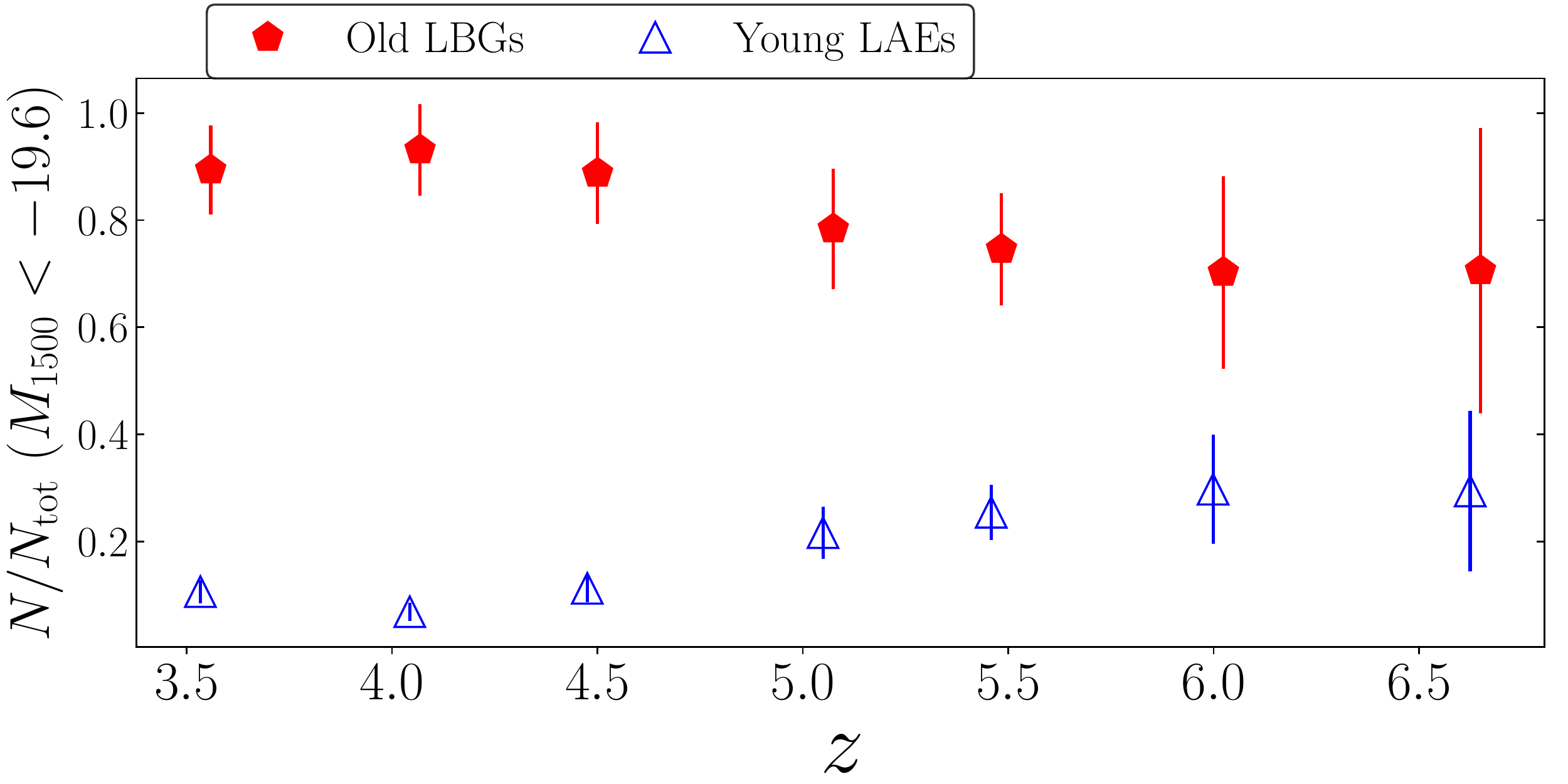}
   \caption{Amount of sources of each class respect to the total number of sources at each $z$ bin. Only objects brighter than \mbox{$M_{1500}=-19.6$} have been considered, corresponding with the approximated 90\% completeness at the highest $z$ bin. The sum of \mbox{no-Ly$\alpha$} LBGs and DSP \mbox{LAE-LBGs} is represented by the red pentagons. Pure LAEs and SSP \mbox{LAE-LBGs} are represented by the blue empty triangles. The trend of these two groups is consistent with an evolutionary scenario between them.}
    \label{fig:relative_numbers}
\end{figure}

Knowing that the Ly$\alpha$ emission quickly decays within the first few Myr, we would expect to see a small fraction of objects presenting strong Ly$\alpha$ line emission among the \mbox{LAE-LBGs} population, which actually matches the EW distribution of this sample presented in \citet{ArrabalHaro2018}. This is also the reason why we find only 404, out of 1,434, continuum sources with emission line. Indeed, we have detected 1,030 sources with no emission line, which would correspond to evolved galaxies with no significant young starburst at present. This represents the most common state of this type of \mbox{high-$z$} galaxies, with the strong Ly$\alpha$ line emission being a recurrent and transiting episode in their lives. Nonetheless, it is also possible that in some sources, especially beyond \mbox{$z\sim5$}, the Ly$\alpha$ photons are destroyed by scattering through a dense neutral medium as discussed in \citet{Hayes2010}. Furthermore, a closer look to the Ly$\alpha$ EW of the LAEs (see Fig.~\ref{fig:EW_dist}) reveals no relation between that and the need of any extra SP for the \mbox{LAE-LBGs}, suggesting that the requirement of DSP models for these \mbox{high-$z$} galaxies is given not only by their Ly$\alpha$ emission but by the relation between this and their emission at longer wavelengths, as also shown in Fig.~\ref{fig:DSP_LAEs_LBGs}, being the Ly$\alpha$ EW an indicator of the age in the SSP models, or of the relative strength of the young SP respect to the old one in the DSP models. On the other side, pure LAEs do show larger Ly$\alpha$ EWs, as expected from their observational definition. Additionally, the relative number of young LAEs increases with $z$ (see Fig.~\ref{fig:relative_numbers}), while that of the old LBGs decreases, which also supports the scenario of an evolution from the pure LAE stage to the LBG, with a much larger proportion of sources in the younger stage the higher the redshift, decreasing as more evolved galaxies form and accumulate as we move to lower redshifts. This behaviour is also consistent with the study of the SFR density (SFRD) carried out by \citet{Sobral2018}, who found an increasing trend with $z$ of the \mbox{$\mathrm{SFRD_{Ly\alpha}/SFRD_{UV}}$} ratio at \mbox{$z=2$-6}.

\subsection{SMFs and SMDs evolution}
\label{sec:SMF_disc}
The obtained SMFs shown in Fig.~\ref{fig:SMFs} are in good agreement with previous estimations from similar studies where the stellar masses were derived through SED fitting. Note that the calculated SMFs are mostly driven by the LBGs population as the bulk of pure LAEs are found in a $M_{\mathrm{star}}$ range out of our completeness, as can be appreciated in Figs.~\ref{fig:mass_dist} and \ref{fig:SFR-M_cigale}. A \mbox{non-negligible} discrepancy can be noticed respect to other works where the UV luminosity was used to derive the $M_{\mathrm{star}}$ through the estimation of a tight mass-to-light ratio \citep{Stark2009, Gonzalez2011, Stefanon2017}. This discrepancy between SED-fitted and \mbox{$M_{\mathrm{UV}}$-derived} $M_{\mathrm{star}}$ could appear due to small differences in the calculated $M/L$ relation, as discussed in \citet{Grazian2015}. The SMF \mbox{best-fitting} Schechter parameters presented in Sec.~\ref{sec:SMF_res} show a decreasing of the SMF with $z$. We find that the characteristic stellar mass $M^{*}$ shifts towards higher masses with cosmic time, as also found by \citet{Grazian2015}. In particular, we measure \mbox{$\log(M_{\mathrm{star}}/M_{\odot})=11.06^{+0.33}_{-0.27}$}, \mbox{$10.78^{+0.53}_{-0.07}$} and \mbox{$10.51^{+0.08}_{-0.03}$} at \mbox{$z\sim4$}, 5 and 6, respectively. No significant change is found for the low mass slope, from \mbox{$\alpha=-1.72^{+0.24}_{-0.14}$} at \mbox{$z\sim4$} to \mbox{$\alpha=-1.76^{+0.19}_{-0.26}$} at \mbox{$z\sim5$} with a slight increase to \mbox{$-1.49^{+0.22}_{-0.21}$} at \mbox{$z\sim6$}. However, the large uncertainty of this last measurement makes it difficult to measure a robust evolution of the $\alpha$ slope within our redshift range. In any case, our $\alpha$ values are much steeper than those found at low redshift, in agreement with previous estimations \citep[\textit{e.g},][]{Santini2012, Duncan2014, Grazian2015, Song2016, Davidzon2017, Stefanon2017}.

The characterisation of the SMF has allowed us to estimate the SMD at each redshift through the integration of the SMF. The results shown in Fig.~\ref{fig:SMD} are in good agreement with the general evolution of this magnitude with cosmic time presented in previous works. Apart from the large errors of the SMD estimation at \mbox{$z\sim5$-6} product of the already discussed uncertainty of the SMF low mass slope calculation at these redshifts, it is worth noticing that the decrease of SMD between \mbox{$z\sim4$-5} is softer than that between \mbox{$z\sim5$-6}, suggesting that the SMD at \mbox{$z\sim5$} obtained in our field could be larger than expected following the general \mbox{SMD-$z$} trend at \mbox{$z=4$-6}. This could be linked to the presence of a reported \mbox{$z\sim5.2$} overdensity in the GOODS-N field \citep{Walter2012, ArrabalHaro2018}.

\subsection{SFR-\texorpdfstring{$M_{\mathrm{star}}$}{} relation and stellar mass growth implications}
\label{sec:SFR-M_disc}
Several authors have previously studied the \mbox{SFR-$M_{\mathrm{star}}$} relation \citep{Stark2009, Gonzalez2010, Papovich2011, Salmon2015} also finding the same lack of evolution on its slope between \mbox{$4\lesssim z\lesssim6$} (see Figs.~\ref{fig:SFR-M} and \ref{fig:SFR-M_cigale}). The low scatter of this relation have suggested that galaxies at this epoch form stars in a larger rate the more massive they are. This supported the hypothesis of a constant pristine gas income over the evolution of these \mbox{high-$z$} galaxies, leaving violent starburst episodes due to mergers or instabilities in a secondary role in the stellar mass growth of these galaxies. However, this stochastic events still likely alter the smooth increase of the $M_{\mathrm{star}}$, as also suggested in \citet{Gonzalez2010} and \citet{Papovich2011}. Our study shows that the assumption of bursty SFHs driving the growth of galaxies at \mbox{$4\lesssim z\lesssim6$} through episodic SF processes is also consistent with the presence of a tight \mbox{SFR-$M_{\mathrm{star}}$} main sequence. It should be noticed that the SFRs derived in \citet{ArrabalHaro2018} from the $L_{1500}$ are the SFR averaged over the last \mbox{30-100 Myr} it takes to the UV luminosity to change after SFR variations \citep[\textit{e.g.},][]{Salim2009,OtiFloranes2010,Salmon2015}. In this way, galaxies which are brighter in the UV do show higher SFRs according to the smooth growth scenario suggested by the \mbox{SFR-$M_{\mathrm{star}}$} main sequence. Nevertheless, these UV brighter and more massive sources are not necessarily presenting the strongest Ly$\alpha$ emission lines, indicating a current (\mbox{$\lesssim25$ Myr}) SF. Additionally, we find a majority of sources (1,030) with undetected Ly$\alpha$ line in our SEDs. These galaxies could still have a smooth and relatively slow star formation component, presenting faint Ly$\alpha$ emission which is not detected in the photometry, with those showing high Ly$\alpha$ EWs suffering a stochastic episode of SF on top of that. In fact, studying the \mbox{SFR-$M_{\mathrm{star}}$} main sequence derived from the \mbox{best-fitting} SSP models (Fig.~\ref{fig:SFR-M_cigale}) we find that both pure LAEs and \mbox{LAE-LBGs} are placed above the mean main sequence, as also found in \citet{Santos2020}, indicating that they are indeed experimenting a recent star-forming episode. 

\section{Conclusions.}
\label{sec:Conclusions}
We have used {\sc cigale} to model the sample of \mbox{high-$z$} LAEs and LBGs selected from the SHARDS survey in \citet{ArrabalHaro2018}, consisting of 1,558 sources at \mbox{$3.4<z<6.8$} in the GOODS-N field. Special attention is given to the differences between the three different subfamilies observationally defined in terms of their Ly$\alpha$ line and UV continuum emission. Single and double stellar population models are used to fit every SED, making use of a Bayesian information criterion calibration to decide in which situations an extra stellar population is needed. With the stellar masses derived from the models, we have studied the SMF, SMD and \mbox{SFR-$M_{\mathrm{star}}$} relation at each $z$, as well as the evolution of the fraction of sources from each subclass. The main conclusions are the following:

\begin{enumerate}
\item The majority (\mbox{$\sim79\%$}) of our \mbox{high-$z$} LAEs and LBGs are well explained by a single stellar population. The cases better described with a secondary stellar population are still strongly dominated by the older population in terms of $M_{\mathrm{star}}$. However, the young stellar population is essential in terms of luminosity to properly fit the Ly$\alpha$ and rest-frame UV emission of these SEDs.

\item The relative amount of objects from each of the subfamilies that need an additional stellar population is not the same. We find that the \mbox{LAE-LBGs}  require double stellar population models in \mbox{$\sim$33\%} of the cases, in comparison with the \mbox{$\sim$16\%} and \mbox{$\sim$22\%} found for the \mbox{no-Ly$\alpha$} LBGs and pure LAEs, respectively. The need of two populations in a significant fraction of the LAEs is due to the presence of a strong Ly$\alpha$ emission line combined with a bright continuum at the longest sampled wavelengths (IRAC) that cannot be simultaneously fitted well by a single stellar population, as in, \textit{e.g.}, \citet{RodriguezEspinosa2014}.

\item Pure LAEs can be tipically understood as very young and low mass galaxies with a median age of \mbox{$\sim26$ Myr} and a median $M_{\mathrm{star}}$ of \mbox{$\sim5\times 10^{8}\ M_{\odot}$}, presenting high Ly$\alpha$ EWs and experimenting one of their first star-forming episodes. The increasing fraction of these objects with $z$ in our sample, consistent with \citet{Sobral2018b}, supports the hypothesis of these pure LAEs typically being an initial and transitional stage on the evolution of \mbox{high-$z$} sources.

\item \mbox{LAE-LBGs} can be split into two subgroups differentiated in age and stellar mass properties. Single stellar population \mbox{LAE-LBGs} seem to be very young (median age of \mbox{$\sim27$ Myr)} but slightly more massive on average than the pure LAEs (median \mbox{$M_{\mathrm{star}}\sim10^{9}\ M_{\odot}$}). The similarities with the pure LAEs subclass suggests that these are members of the same kind of young galaxies, but with different SFRs and stellar masses. The relative number of young LAEs follows the same trend with $z$ than the pure LAEs one, supporting the idea of them being the same kind of galaxies.

\item Dual stellar population LAEs are fitted by older (hundreds of Myr) and more massive models (\mbox{$M_{\mathrm{star}}\sim10^{10}\ M_{\odot}$}) featuring a young and much less massive population causing the bulk of the Ly$\alpha$ emission. According to this, double stellar population LAEs seem to be galaxies more massive and evolved (at this $z$), undergoing an episodic star-forming episode.

\item \mbox{No-Ly$\alpha$} LBGs are the most difficult subclass to model, as they do not show emission patterns that strongly help to constrain their ages in the rest-frame wavelength range sampled in this work at these redshifts. This creates a degeneracy in the combinations of stellar populations that could lead to a good fit of their SEDs with negligible variations in the $\chi^{2}$. Moreover, some of these galaxies could actually host a very young stellar population whose Ly$\alpha$ line is not detected because of resonant scattering and dust extinction. Furthermore, the use of the Bayesian information criterion could be biasing the calculated ages for this family towards younger values, so we only aim to model these sources in a general way, being aware that the description of the class may not match all the individual cases.

\item With the caveats just mentioned, the results derived from {\sc cigale} show that \mbox{no-Ly$\alpha$} LBGs lack a really young SP (\mbox{$\lesssim25$ Myr}). Furthermore, the absence of strong Ly$\alpha$ emission indicates that these sources are not in a current strong star-forming episode (or have extremely low escape fractions). However, it is possible that these galaxies present a fairly smooth star formation, producing faint Ly$\alpha$ lines which are not detected in the photometry. They are older and much more massive than pure LAEs or single stellar population \mbox{LAE-LBGs}, with a median $M_{\mathrm{star}}$ of \mbox{$\sim3.5\times 10^{9}\ M_{\odot}$}. These results suggest that \mbox{no-Ly$\alpha$} LBGs are a more evolved stage of \mbox{high-$z$} galaxies that have been forming stars for a longer time, developing larger stellar masses and presenting brighter continuum emission at longest wavelengths, because of the old stars. The evolution of the fraction of these objects with $z$ also supports the idea of \mbox{no-Ly$\alpha$} LBGs being more evolved star-forming sources, the more common the lower the $z$ is.

\item We report a decreasing evolution of the characteristic stellar mass of the SMFs with $z$, as in \textit{e.g.}, \citet{Grazian2015}, finding \mbox{$\log(M^{*}/M_{\odot})=11.06^{+0.33}_{-0.27}$}, \mbox{$10.78^{+0.53}_{-0.07}$} and \mbox{$10.51^{+0.08}_{-0.03}$} at \mbox{$z\sim4$}, 5 and 6, respectively. The low mass slopes found are steeper than those typically found at low redshift. No significant evolution is found between \mbox{$z=4-$5}, with a small increase at \mbox{$z\sim6$} (\mbox{$\alpha=-1.72^{+0.24}_{-0.14}$}, \mbox{$-1.76^{+0.19}_{-0.26}$} and \mbox{$-1.49^{+0.22}_{-0.21}$} at \mbox{$z\sim4$}, 5 and 6, respectively). However, the $\alpha$ estimated at \mbox{$z\sim5$-6} has to be carefully considered, as we do not have much information covering the $M_{\mathrm{star}}$ region corresponding to the potential term of the Schechter SMF at these redshifts.

\item The SMD is estimated by integration of the SMF at each redshift. Our results are in agreement with the \mbox{SMD-$z$} trend reported at these redshifts by previous authors \citep{Labbe2010, Gonzalez2011, Lee2012, Duncan2014, Grazian2015, Song2016}. The SMD obtained at \mbox{$z\sim5$}, although consistent with the general trend, is slightly larger than expected if we follow the mean slope of the \mbox{SMD-$z$} relation at \mbox{high-$z$}, which could be linked to the presence of a previously reported \mbox{$z\sim5.2$} overdensity in GOODS-N \citep{Walter2012, ArrabalHaro2018}. Additional research is incoming to further characterise this overdensity.

\item The slope values found for the \mbox{$\mathrm{SFR}\propto M_{\mathrm{star}}^{\beta}$} relation are \mbox{$\beta=0.48^{+0.07}_{-0.10}$}, \mbox{$0.46^{+0.12}_{-0.10}$}, \mbox{$0.51^{+0.26}_{-0.19}$} at \mbox{$z\sim4$}, 5 and 6, respectively, for the \mbox{UV-derived} SFRs and \mbox{$\beta=0.83^{+0.09}_{-0.06}$}, \mbox{$0.79^{+0.10}_{-0.11}$}, \mbox{$0.82^{+0.15}_{-0.16}$} at \mbox{$z\sim4$}, 5 and 6, respectively, for the \mbox{model-derived} SFRs, both of them consistent with little to no redshift evolution of that slope within that redshift range, in agreement with previous works \citep[\textit{e.g.},][]{Stark2009, Gonzalez2010, Papovich2011, Salmon2015}. The existence of such tight relation between these two magnitudes and its invariability within this $z$ range point to the hypothesis of a smooth pristine gas infall as the main mechanism responsible of the mass growth of these galaxies along their lives, as suggested before. Nevertheless, the \mbox{best-fitting} \mbox{burst-like} SFHs used in this work also produce an equally tight main sequence. This, joined to the fact that LAEs appear above the \mbox{SFR-$M_{\mathrm{star}}$} main sequence, supports the existence of stochastic star-forming events due to mergers and other instabilities that can also be responsible of the stellar mass growth in \mbox{high-$z$} galaxies.
\end{enumerate}

\section*{Acknowledgements}
%Try to keep it short.
We want to acknowledge support from the Spanish Ministry of Economy and Competitiveness (MINECO) under grants \mbox{AYA2015-70498-C2-1-R}, \mbox{AYA2013-47742-C4-2-P} and \mbox{AYA2016-79724-C4-2-P}. Based on observations made with the Gran Telescopio Canarias (GTC), installed in the Spanish Observatorio del Roque de los Muchachos of the Instituto de Astrofísica de Canarias, in the island of La Palma.

We would like to thank Dr. Jairo Méndez Abreu for his help on the use and calibration of the Bayesian information criterion and Dr. J. Miguel Mass Hesse for several discussions about stellar population models. Special thanks to the anonymous referee for very useful and detailed comments which have really improved this work.

%%%%%%%%%%%%%%%%%%%%%%%%%%%%%%%%%%%%%%%%%%%%%%%%%%

%%%%%%%%%%%%%%%%%%%% REFERENCES %%%%%%%%%%%%%%%%%%

% The best way to enter references is to use BibTeX:

\bibliographystyle{apj_modified}
\bibliography{biblio}

% Alternatively you could enter them by hand, like this:
% This method is tedious and prone to error if you have lots of references
%\begin{thebibliography}{99}

%\end{thebibliography}

%%%%%%%%%%%%%%%%%%%%%%%%%%%%%%%%%%%%%%%%%%%%%%%%%%
\clearpage

\appendix
\section{Monte Carlo Schechter function fit to the SMF}
\label{sec:appendix}
\begin{figure}
	\includegraphics[width=\columnwidth]{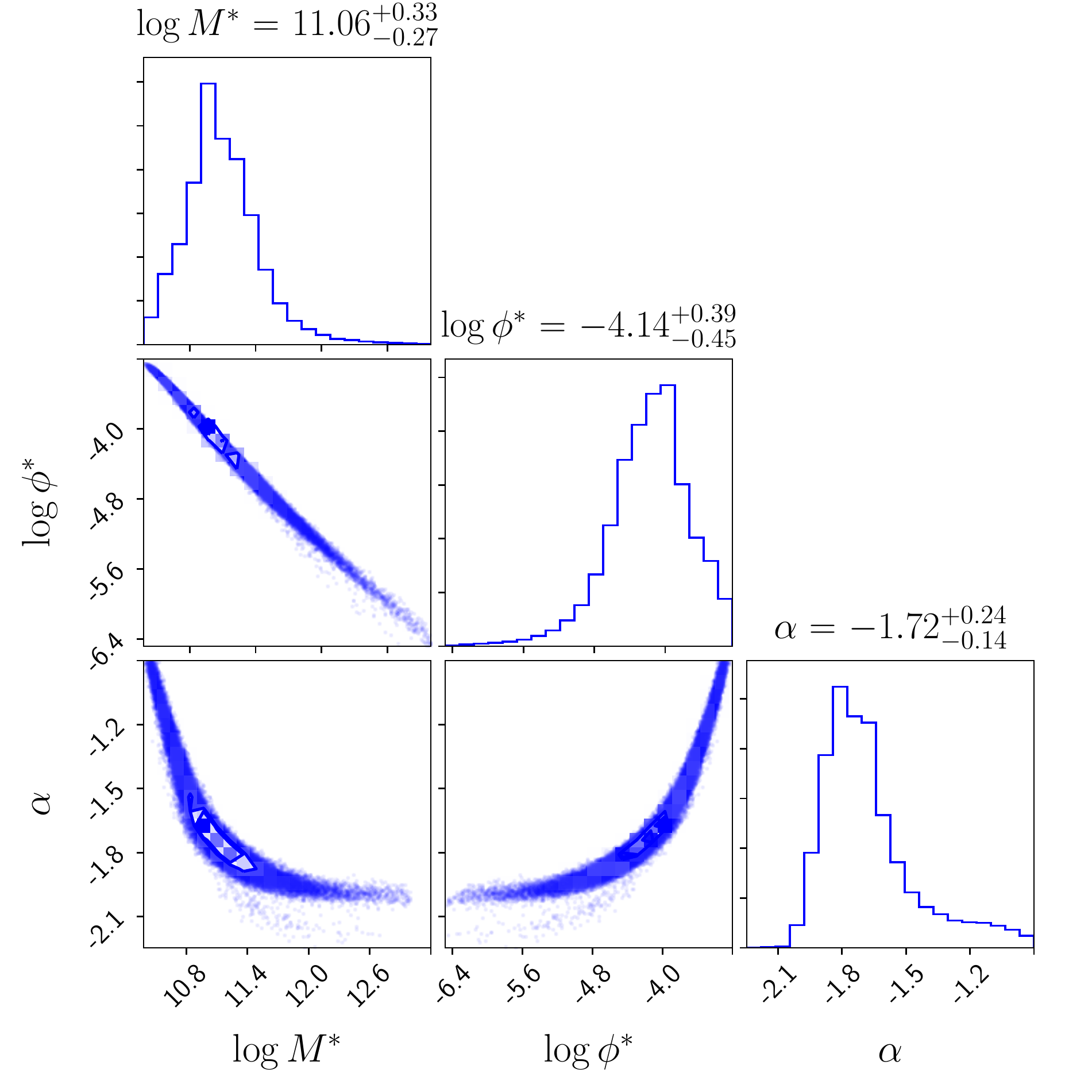}
   \caption{Confidence intervals on the Schechter parameters from the Monte Carlo fitting at \mbox{$z\sim4$}. The inner and outer contours correspond to the $1\sigma$ and $3\sigma$ significance, respectively.}
    \label{fig:contours_4}
\end{figure}

\begin{figure}
	\includegraphics[width=\columnwidth]{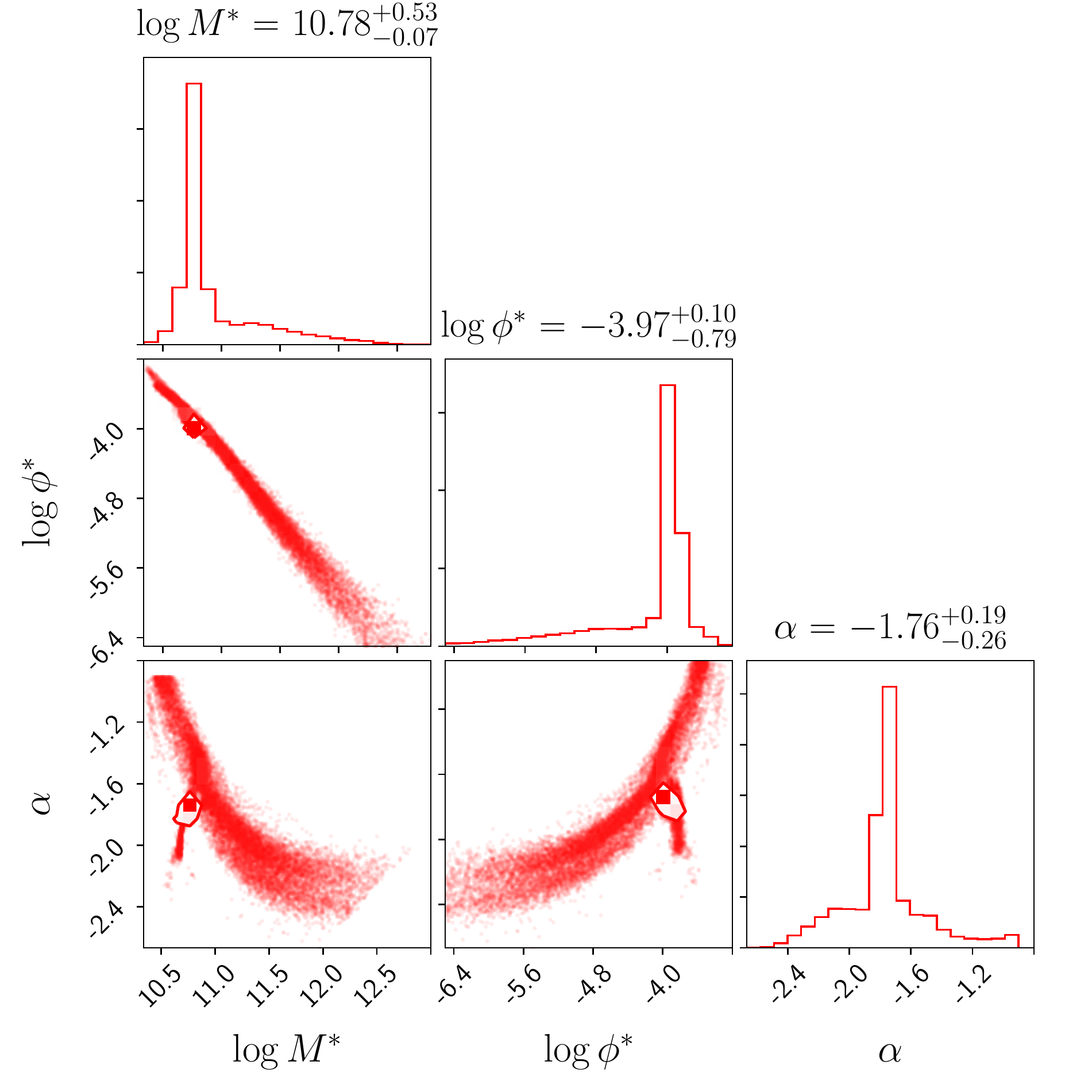}
   \caption{Confidence intervals on the Schechter parameters from the Monte Carlo fitting at \mbox{$z\sim5$}. The inner and outer contours correspond to the $1\sigma$ and $3\sigma$ significance, respectively.}
    \label{fig:contours_5}
\end{figure}

\begin{figure}
	\includegraphics[width=\columnwidth]{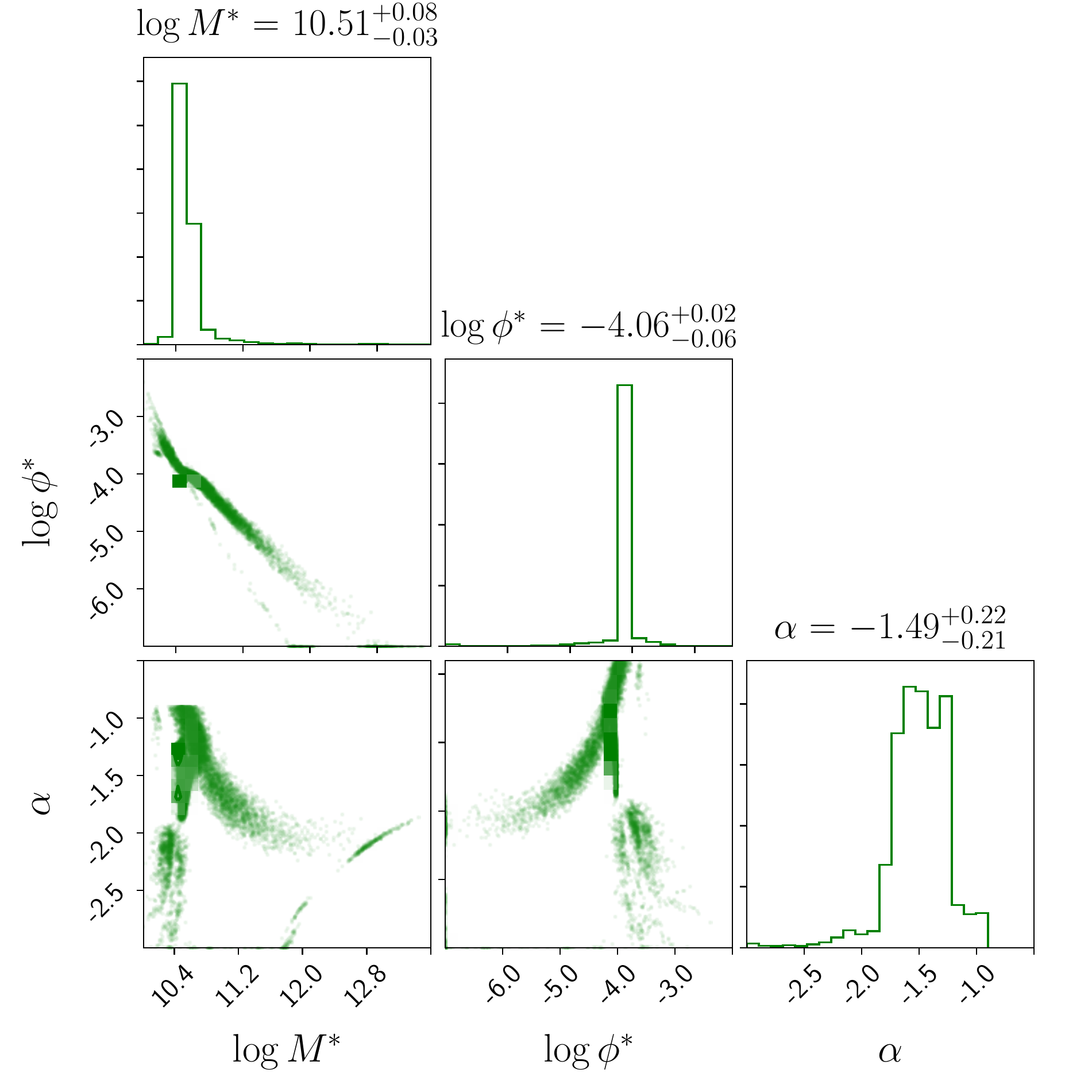}
   \caption{Confidence intervals on the Schechter parameters from the Monte Carlo fitting at \mbox{$z\sim6$}. The inner and outer contours correspond to the $1\sigma$ and $3\sigma$ significance, respectively.}
    \label{fig:contours_6}
\end{figure}

In this appendix we show the confidence intervals of the Schechter parameters obtained using Monte Carlo simulations to fit the SMF. The perturbations in the measurements are implemented injecting a Gaussian noise to each point consistent with its own Poissonian error. Additionally, the $M_{\mathrm{star}}$ bins are also perturbed, both in size and centre value. In particular, the bin size is perturbed between \mbox{0.2-0.3 dex} in intervals of 0.025 dex. The central $M_{\mathrm{star}}$ value of each bin is as well shifted 0.1 dex in intervals of 0.025 dex. At \mbox{$z\sim6$}, using the SMF points up to the estimated stellar mass 90\% completeness limit results in the inclusion of actually incomplete points close to the characteristic $M^{*}$ knee for several $M_{\mathrm{star}}$ bin perturbations, obtaining positive values for the low mass slope. To solve this issue, only strictly increasing SMF points are considered for the Schechter function fit as we move to lower stellar masses up to our $M_{\mathrm{star},\mathrm{lim}}$, as further points are considered incomplete. The $\alpha$ slope is also constrained to \mbox{$-3.0<\alpha<-0.9$} for the fit. We are aware that this approach could bias our \mbox{$z\sim6$} low mass slope estimation towards steeper values, as warned in the text. The significance contours of the three Schechter parameters at \mbox{$z\sim4$}, 5 and 6 are shown in Figs.~\ref{fig:contours_4}, \ref{fig:contours_5} and \ref{fig:contours_6}, respectively.

%\FloatBarrier

%\FloatBarrier

% Don't change these lines
\bsp	% typesetting comment
\label{lastpage}
\end{document}